\documentclass[a4paper]{amsart}

\usepackage{amsfonts,amssymb,amsmath,amsthm,gensymb,color,epic,eepic,float,fullpage}
\usepackage[mathscr]{euscript}
\usepackage[T1]{fontenc}
\usepackage{mathtools}
\allowdisplaybreaks 

			\newtheorem*{myTheorem}{Theorem}
			
			\renewcommand{\r}[1]{(\ref{#1})}
			\newcommand{\s}[1]{\S\ref{#1}}
			\renewcommand{\mod}[2]{\text{\tiny{mod}}_{#1}(#2)}
			\newcommand{\res}[2]{\text{res}_{#1=#2}}

			\newcommand*{\gfrac}[2]{\genfrac{}{}{0pt}{}{#1}{#2}}

			\newcommand*{\llangle}{\langle\kern-2\nulldelimiterspace\langle}
			\newcommand*{\rrangle}{\rangle\kern-2\nulldelimiterspace\rangle}

			\newcommand{\bvec}[2]{\| #1, #2 \rrangle}
			\newcommand{\dbvec}[2]{\llangle #1, #2 \|}

\def\wh{\widehat}
\def\b{\bar}

\def\leq{\leqslant}

\def\dfn{=}

\def\i{i}

\def\ivan{\overrightarrow{\Delta}}
\def\dvan{\overleftarrow{\Delta}}

\def\l{\lambda}
\def\m{\mu}

\def\li{\b\lambda_{\mbox{\rm\tiny I}}}
\def\lii{\b\lambda_{\mbox{\rm\tiny II}}}
\def\mi{\b\mu_{\mbox{\rm\tiny I}}}
\def\mii{\b\mu_{\mbox{\rm\tiny II}}}
\def\lc{\lambda^{C}}
\def\blc{\b\lambda^{C}}
\def\wlc{\wh\lambda^{C}}
\def\lci{\b\lambda^{C}_{\mbox{\rm\tiny I}}}
\def\lcii{\b\lambda^{C}_{\mbox{\rm\tiny II}}}
\def\lb{\lambda^{B}}
\def\blb{\b\lambda^{B}}
\def\wlb{\wh\lambda^{B}}
\def\lbi{\b\lambda^{B}_{\mbox{\rm\tiny I}}}
\def\lbii{\b\lambda^{B}_{\mbox{\rm\tiny II}}}
\def\mc{\mu^{C}}
\def\bmc{\b\mu^{C}}
\def\wmc{\wh\mu^{C}}
\def\mci{\b\mu^{C}_{\mbox{\rm\tiny I}}}
\def\mcii{\b\mu^{C}_{\mbox{\rm\tiny II}}}
\def\mb{\mu^{B}}
\def\bmb{\b\mu^{B}}
\def\wmb{\wh\mu^{B}}
\def\mbi{\b\mu^{B}_{\mbox{\rm\tiny I}}}
\def\mbii{\b\mu^{B}_{\mbox{\rm\tiny II}}}

\def\ie{{\it i.e.}\/}

\begin{document}

\title{Multiple integral formulae for the scalar product of on-shell and off-shell Bethe vectors in $SU(3)$-invariant models}

\author{M Wheeler}

\address{Laboratoire de Physique Th\'eorique et Hautes Energies, CNRS UMR 7589 and Universit\'e Pierre et Marie Curie (Paris 6), 4 place Jussieu, 75252 Paris cedex 05, France}
\email{mwheeler@lpthe.jussieu.fr}

\keywords{$SU(2)$ and $SU(3)$-invariant models. Nested Bethe Ansatz. Scalar products.}

\begin{abstract}
We study the scalar product $\mathcal{S}_{\ell,m}$ between an on-shell and an off-shell Bethe state in models with $SU(3)$-invariance, where $\ell$ and $m$ denote the cardinalities of the two sets of Bethe roots. We construct recursion relations relating $\mathcal{S}_{\ell,m}$ to scalar products of smaller dimension, namely $\mathcal{S}_{\ell-1,m}$ and $\mathcal{S}_{\ell,m-1}$. Solving these recursion relations we obtain new multiple integral expressions for $\mathcal{S}_{\ell,m}$, whose integrands are $(\ell+m) \times (\ell+m)$ determinants, and closely related to the Slavnov determinant expression for the $SU(2)$ scalar product.
\end{abstract} 

\maketitle

\setcounter{section}{0}

\section{Introduction}
\label{s:intro}

It has been known for some time that scalar products play an important role in quantum integrable models solvable by the Bethe Ansatz, since they are quantities which appear naturally in the calculation of correlation functions within these models. Scalar products have been well studied in models based on the $SU(2)$-invariant $R$-matrix. The term itself originated in the works \cite{kor,ik} by Korepin and Izergin, where the authors obtained a sum expression for the scalar product between two off-shell Bethe vectors in a generic $SU(2)$-invariant model. Prior to this, a determinant formula for the scalar product between two on-shell Bethe vectors (the norm-squared) had been conjectured by Gaudin \cite{gau}, and was proved by Korepin in \cite{kor}. Interpolating between the off-shell/off-shell scalar product and the Gaudin norm, the scalar product between an {\it on-shell} and {\it off-shell} Bethe vector was evaluated in determinant form\footnote{In the case of the on-shell/off-shell scalar product in the XXX spin-1/2 chain, alternative determinant expressions have also been obtained in \cite{km,fw3}.} by Slavnov \cite{sla}. The latter representation has been of great use in the algebraic Bethe Ansatz approach to correlation functions of the XXX and XXZ models (see \cite{kmt}, the review article \cite{kmst}, and references therein). More recently, it has also appeared in the calculation of 3-point functions of local gauge-invariant single-trace operators in the scalar sector of planar $\mathcal{N}=4$ supersymmetric Yang-Mills \cite{egsv,fod,kos1,kos2}.

A natural problem is to extend these earlier works to models based on higher rank algebras. In the recent work of Belliard, Pakuliak, Ragoucy and Slavnov in \cite{bprs1,bprs2,bprs3,bprs4,bprs5}, the authors have made significant progress in the study of scalar products and correlation functions in quantum integrable models with an $SU(3)$-invariant $R$-matrix. A key result in this series of papers, in \cite{bprs2}, is a determinant representation for the scalar product between an 
on-shell Bethe vector (eigenvector of the transfer matrix) and a twisted on-shell Bethe vector (eigenvector of the twisted transfer matrix). This result generalizes the determinant expression for the on-shell/on-shell scalar product in an $SU(3)$-invariant model, obtained by Reshetikhin in \cite{res}, and was crucial to obtaining single determinant expressions for particular form factors in \cite{bprs4}.

Other important results in this series, in \cite{bprs3}, are summation expressions for the action of arbitrarily many monodromy matrix elements on off-shell Bethe vectors in $GL(3)$-invariant models. The authors were able to demonstrate the (highly non-trivial) fact that acting with any monodromy matrix element on an off-shell Bethe vector produces a sum over off-shell Bethe vectors, with appropriate coefficients. This result lays the foundation for calculating any correlation function in these models as a sum over on-shell/off-shell scalar products. On the other hand, such an expression is of limited use without an appropriate generalization of the Slavnov formula to higher rank. To date, {\it no single determinant expression for the on-shell/off-shell scalar product at higher rank is known.} Nevertheless, the current paper presents (what we hope will be) a computationally tractable formula for the on-shell/off-shell scalar product in $SU(3)$-invariant models, as a multiple integral over a determinant which generalizes the $SU(2)$ Slavnov determinant. 

The main idea in this work is to use the inherent left/right asymmetry in the on-shell/off-shell scalar product to obtain information about it. More specifically, if $|\{\lc\},\{\mc\}\rangle$ is an off-shell Bethe vector (parametrized by two sets of free variables $\{\lc\}$ and $\{\mc\}$ with respective cardinalities $\ell$ and $m$) in the Hilbert space $\mathcal{H}$ of an $SU(3)$-invariant model, and $\langle \{\mb\}, \{\lb\}|$ is an 
on-shell Bethe vector (parametrized by two sets of Bethe roots $\{\lb\}$ and $\{\mb\}$, also with cardinalities $\ell$ and $m$) in the dual space $\mathcal{H}^{*}$, it is natural to consider the quantity
\begin{align*}
\mathcal{S}(z) 
= 
\langle \{\mb\}, \{\lb\} | \mathcal{T}(z) | \{\lc\},\{\mc\} \rangle
\end{align*}
where $\mathcal{T}(z)$ denotes the transfer matrix of the model. There are two ways to calculate 
$\mathcal{S}(z)$. One can act with $\mathcal{T}(z)$ to the left, which produces the on-shell/off-shell scalar product up to a multiplicative term, since $\langle \{\mb\}, \{\lb\}|$ is an eigenstate of the transfer matrix. On the other hand, acting with $\mathcal{T}(z)$ to the right, one can use the results of \cite{bprs3} to write $\mathcal{T}(z) | \{\lc\},\{\mc\} \rangle$ as a sum over off-shell states. Equating the two approaches, and sending $z$ to the value of one of the Bethe roots (either $z \rightarrow \lb_i$ for some 
$1\leq i \leq \ell$, or $z \rightarrow \mb_j$ for some $1\leq j \leq m$), we obtain recursion relations expressing $\mathcal{S}_{\ell,m}$ as a sum over on-shell/off-shell scalar products of smaller size, $\mathcal{S}_{\ell-1,m}$ and $\mathcal{S}_{\ell,m-1}$. Once one has obtained such recursion relations, it is straightforward to prove that the multiple integral expressions we have found are indeed solutions. This method of constructing a functional relation for a scalar product, and the idea that it should have a multiple integral solution, has appeared previously in the context of the $SU(2)$-invariant XXX spin-chain in \cite{gal1}.\footnote{We thank W~Galleas for bringing this to our attention.}

The paper is organized as follows: in \s{s:not-def} we give some preliminary remarks regarding our notation, define what is meant by a generalized $SU(3)$-invariant model, write down expressions for the Bethe vectors (obtained via the nested Bethe Ansatz \cite{kr,res,br}) and define their scalar product. In \s{s:results} we review the determinant expression for on-shell/off-shell $SU(2)$ scalar product, obtained in \cite{sla}, and write down our multiple integral formulae for the same object in the $SU(3)$ case. \s{s:review} and \s{s:recursion} serve as an extended proof of these formulae. In \s{s:review} we review some of the results obtained in \cite{bprs3}, in particular the expression for $\mathcal{T}(z) | \{\lc\},\{\mc\} \rangle$ as a sum over off-shell states. In \s{s:recursion}, proceeding along the lines described in the previous paragraph, we derive recursion relations for the on-shell/off-shell $SU(3)$ scalar product and prove that the multiple integral formulae are solutions. Finally, as a consistency test, in \s{s:cases} we consider the limiting cases 
$\{\lb\} \rightarrow \infty$ and $\{\mb\} \rightarrow \infty$ of the multiple integral expressions. We find that in these limits the formulae simplify greatly, and factorize into a product of two determinants, reproducing the main result of \cite{whe,fw2}. We conclude in \s{s:discuss}. 

\section{Notation, nested Bethe Ansatz and definition of the scalar product}
\label{s:not-def}

\subsection{Notation related to sets}

Throughout the paper, we indicate a set of variables by placing a bar over a letter. For example
\begin{align*}
\b{x}_m = \{x_1,\dots,x_m\},
\quad\quad
\b{y}_n = \{y_1,\dots,y_n\},
\end{align*}
where the use of the subscript indicates the cardinality of the set, and that the elements are labelled with subscripts ranging from 1 up to the cardinality. When the cardinality and labelling is clear from context, we may omit the subscript, writing for example $\b{x}_m \equiv \b{x}$.

In the case where a single variable is excluded from a set, we indicate this by placing a circumflex over the letter. For example
\begin{align*}
\wh{x}_{m,i} = \{x_1, \dots, x_{i-1}, x_{i+1}, \dots, x_m\},
\quad\quad
\wh{y}_{n,j} = \{y_1,\dots, y_{j-1}, y_{j+1}, \dots, y_n\},
\end{align*}
where the first subscript indicates the cardinality, and the second subscript indicates the variable to be omitted. Again, when the cardinality is clear from context, we simplify the notation by writing for example $\wh{x}_{m,i} \equiv \wh{x}_i$.

In place of the usual symbols for union and exclusion, $\cup$ and $\backslash$, we shall instead use $\oplus$ and $\ominus$. For example, we write
\begin{align*}
\b{x}_m \oplus \b{y}_n
=
\{x_1,\dots,x_m\} \oplus \{y_1,\dots,y_n\}
=
\{x_1,\dots,x_m,y_1,\dots,y_n\}
\end{align*}
for the union of two sets, and
\begin{align*}
(\b{x}_m \oplus \b{y}_n) \ominus \b{y}_n
= 
\{x_1,\dots,x_m,y_1,\dots,y_n\} \ominus \{y_1,\dots,y_n\}
=
\{x_1,\dots,x_m\}
\end{align*}
for the exclusion of a subset.

\subsection{Commonly used functions}

Three types of rational functions are used frequently throughout the paper:
\begin{align*}
f(x,y) = \frac{x-y+1}{x-y},
\quad\quad
g(x,y) = \frac{1}{x-y},
\quad\quad
h(x,y) = x-y+1
\end{align*}
When one of these functions takes a set as an argument, this means taking a product of that function over all values in the set:
\begin{align*}
f(x,\b{y}_n) 
=
\prod_{j=1}^{n}
f(x,y_j),
\quad\quad
f(\b{x}_m,y)
=
\prod_{i=1}^{m}
f(x_i,y),
\quad\quad
f(\b{x}_m,\b{y}_n) = 
\prod_{i=1}^{m}
\prod_{j=1}^{n} f(x_i,y_j)
\end{align*}
For example, combining all of this notation, if $\b{z}_n \subseteq \b{x}_{\ell} \oplus \b{y}_m$ then
\begin{align*}
f(w,\b{x}_{\ell} \oplus \b{y}_m \ominus \b{z}_n)
=
\frac{
\prod_{i=1}^{\ell} f(w,x_i) \prod_{j=1}^{m} f(w,y_j)
}
{
\prod_{k=1}^{n} f(w,z_k)
} 
\end{align*}
Clearly this formula continues to make sense even if $\b{z}_n$ is not a subset of $\b{x}_{\ell} \oplus \b{y}_m$.

\subsection{$R$-matrices and related definitions}

Let $V_{\alpha}, V_{\beta}$ be two copies of the vector space $\mathbb{C}^3$. The $SU(3)$-invariant $R$-matrix is given by
\begin{align}
R_{\alpha\beta}^{(1)}(\lambda,\mu)
=
\left(
\begin{array}{ccc|ccc|ccc}
f(\lambda,\mu) & 0 & 0 & 0 & 0 & 0 & 0 & 0 & 0 \\
0 & 1 & 0 & g(\lambda,\mu) & 0 & 0 & 0 & 0 & 0 \\
0 & 0 & 1 & 0 & 0 & 0 & g(\lambda,\mu) & 0 & 0 \\ \hline
0 & g(\lambda,\mu) & 0 & 1 & 0 & 0 & 0 & 0 & 0 \\
0 & 0 & 0 & 0 & f(\lambda,\mu) & 0 & 0 & 0 & 0 \\
0 & 0 & 0 & 0 & 0 & 1 & 0 & g(\lambda,\mu) & 0 \\ \hline
0 & 0 & g(\lambda,\mu) & 0 & 0 & 0 & 1 & 0 & 0 \\
0 & 0 & 0 & 0 & 0 & g(\lambda,\mu) & 0 & 1 & 0 \\
0 & 0 & 0 & 0 & 0 & 0 & 0 & 0 & f(\lambda,\mu) 
\end{array}
\right)_{\alpha\beta}
\label{Rmat-su3}
\end{align}
where the subscript indicates that the $R$-matrix is an element of 
${\rm End}(V_{\alpha} \otimes V_{\beta})$. 

Because the nested Bethe Ansatz involves a reduction of the $SU(3)$ eigenvector problem to an $SU(2)$ one, we also need the definition
\begin{align}
R^{(2)}_{\alpha\beta}(\lambda,\mu)
=
\left(
\begin{array}{cc|cc}
f(\lambda,\mu) & 0 & 0 & 0 \\
0 & 1 & g(\lambda,\mu) & 0 \\ \hline
0 & g(\lambda,\mu) & 1 & 0 \\
0 & 0 & 0 & f(\lambda,\mu)
\end{array}
\right)_{\alpha\beta}
\label{Rmat-su32}
\end{align}
which is simply the $SU(2)$-invariant $R$-matrix, where $V_{\alpha},V_{\beta}$ are copies of $\mathbb{C}^2$. 

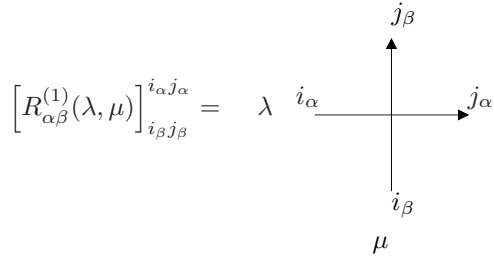
\begin{figure}[H]

\begin{center}
\begin{minipage}{4.3in}

\setlength{\unitlength}{0.0005cm}
\begin{picture}(5000,6000)(3000,16500)

\put(6000,20000)
{$
\Big[R_{\alpha\beta}^{(1)}(\lambda,\mu)\Big]^{i_{\alpha} j_{\alpha}}_{i_{\beta} j_{\beta}}
=
$}


\path(14000,20000)(18000,20000) \put(12500,20000){$\lambda$}
\put(13500,20300){$i_{\alpha}$}
\put(18000,20300){$j_{\alpha}$}
\blacken\path(17750,20125)(17750,19875)(18000,20000)(17750,20125) 


\path(16000,18000)(16000,22000) \put(15500,16500){$\mu$}
\put(16000,17500){$i_{\beta}$}
\put(16000,22500){$j_{\beta}$}
\blacken\path(16125,21750)(15875,21750)(16000,22000)(16125,21750)

\end{picture}

\end{minipage}
\end{center}

\caption{Representing the components of the $R$-matrix as vertices. The indices $i_{\alpha},j_{\alpha} \in \{1,2,3\}$ denote the $(i_{\alpha},j_{\alpha})$-th block of (\ref{Rmat-su3}), while $i_{\beta},j_{\beta} \in \{1,2,3\}$ denote the $(i_{\beta},j_{\beta})$-th component within that block. The black arrows placed on the lines are only to fix unambiguously the orientation of a vertex.}

\label{f:vert1}

\end{figure}

\subsection{Generalized $SU(3)$-invariant models}
\label{ssec:gen-su3}

We consider a generalized $SU(3)$-invariant model\footnote{For more information on what is meant by \textit{generalized $SU(3)$-invariant model}, we refer the reader to \cite{kor2,tar1,tar2}, which define this concept in detail in the $SU(2)$-invariant case.} with the $3\times 3$ monodromy matrix
\begin{align}
T^{(1)}_{\alpha}(\lambda)
=
\left(
\begin{array}{ccc}
T_{11}(\lambda) & T_{12}(\lambda) & T_{13}(\lambda) \\
T_{21}(\lambda) & T_{22}(\lambda) & T_{23}(\lambda) \\
T_{31}(\lambda) & T_{32}(\lambda) & T_{33}(\lambda)
\end{array}
\right)_{\alpha} 
\label{mon-su3}
\end{align}
The entries of (\ref{mon-su3}) are operators which satisfy the Yang-Baxter algebra
\begin{align}
R_{\alpha\beta}^{(1)}(\lambda,\mu)
T_{\alpha}^{(1)}(\lambda)
T_{\beta}^{(1)}(\mu)
=
T_{\beta}^{(1)}(\mu)
T_{\alpha}^{(1)}(\lambda)
R_{\alpha\beta}^{(1)}(\lambda,\mu)
\label{int-su3}
\end{align}
where $R_{\alpha\beta}^{(1)}(\lambda,\mu)$ is the $SU(3)$-invariant $R$-matrix (\ref{Rmat-su3}). The model is solvable by the nested Bethe Ansatz \cite{kr,res,br} if one can find pseudo-vacuum states $|0\rangle,\langle 0|$ which are acted upon by the operators (\ref{mon-su3}) according to the rules 
\begin{align}
T_{ii}(\lambda) |0\rangle = a_i(\lambda) |0\rangle,
\quad
T_{kj}(\lambda) |0\rangle = 0,
\quad
T_{jk}(\lambda) |0\rangle \not= 0
\label{ps-vac3}
\\
\langle 0| T_{ii}(\lambda) = a_i(\lambda) \langle 0|,
\quad
\langle 0| T_{kj}(\lambda) \not= 0,
\quad
\langle 0| T_{jk}(\lambda) = 0
\label{ps-vac4}
\end{align}
valid for all $i \in \{1,2,3\}$ and $1\leq j < k \leq 3$, and where the $a_i(\lambda)$ are rational functions (which we call the pseudo-vacuum eigenfunctions). Let $\mathcal{H}$ denote the Hilbert space generated by the action of operators $T_{jk}(\lambda)$ on $|0\rangle$, and $\mathcal{H}^{*}$ the Hilbert space generated by $T_{kj}(\lambda)$ on $\langle 0|$, for all $1\leq j < k \leq 3$.

The transfer matrix $\mathcal{T}(\lambda)$ is the trace of the monodromy matrix \r{mon-su3} on the space $V_{\alpha}$:
\begin{align}
\mathcal{T}(\lambda) = \sum_{k=1}^{3} T_{kk}(\lambda)
\label{transfer}
\end{align}
and the goal of the nested Bethe Ansatz is to find its eigenvectors and eigenvalues, \ie\ all
$|\Psi\rangle \in \mathcal{H}$ and $\Lambda_{\Psi}(\lambda) \in \mathbb{C}$ such that 
$\mathcal{T}(\lambda) |\Psi\rangle = \Lambda_{\Psi}(\lambda) |\Psi\rangle$.

\subsection{Decomposition of monodromy matrix}

In order to construct the eigenvectors of \r{transfer}, consider a $2 \times 2$ decomposition of the monodromy matrix,\footnote{In this section we will label $B$ and $C$ operators in the opposite sense to how they are typically labelled in the literature. For example, what we have called $C^{(1)}$ and $C^{(2)}$ would usually be denoted $B^{(1)}$ and $B^{(2)}$, and vice versa. We do this for consistency with the labelling of the variables that we use subsequently. Since we never use these operators beyond this section, we hope this will not cause any confusion.} by defining $A^{(1)}(\lambda) = T_{11}(\lambda)$ and
\begin{align}
B_{\beta}^{(1)}(\lambda)
=
\left(
\begin{array}{c}
T_{21}(\lambda) \\
T_{31}(\lambda)
\end{array}
\right)_{\beta},
\quad\quad
C_{\gamma}^{(1)}(\lambda) 
=
\left(
\begin{array}{cc}
T_{12}(\lambda) & T_{13}(\lambda)
\end{array}
\right)_{\gamma},
\quad\quad
D^{(1)}_{\delta}(\lambda)
=
\left(
\begin{array}{cc}
T_{22}(\lambda) & T_{23}(\lambda) \\
T_{32}(\lambda) & T_{33}(\lambda)
\end{array}
\right)_{\delta}
\label{decomposition} 
\end{align}
Here $V_{\beta}, V_{\gamma}, V_{\delta}$ are copies of $\mathbb{C}^2$ and the subscripts on these matrices are used to denote the fact that
\begin{align*} 
B^{(1)}_{\beta}(\lambda) \in V_{\beta}, \quad\quad
C^{(1)}_{\gamma}(\lambda) \in V_{\gamma}^{*},\quad\quad
D^{(1)}_{\delta}(\lambda) \in {\rm End}(V_{\delta})
\end{align*}

\subsection{Generic Bethe vectors (off-shell states)}

Although we will not describe its details, the nested Bethe Ansatz leads one to propose that the eigenvectors of \r{transfer} are of the form 
\begin{align}
|\Psi\rangle
=
|\b{\l}_{\ell}, \b{\m}_m\rangle
=
C^{(1)}_{\alpha_1}(\lambda_1)
\ldots
C^{(1)}_{\alpha_{\ell}}(\lambda_{\ell})
C^{(2)}(\mu_1)
\ldots
C^{(2)}(\mu_m)
|0\rangle \otimes |\Uparrow_{\alpha}\rangle
\label{gen-bethe}
\end{align}
where each $C^{(1)}_{\alpha_i}(\lambda_i)$ is a row vector given by (\ref{decomposition}), and $C^{(2)}(\m)$ denotes the $(1,2)$-th component of the $SU(2)$-invariant monodromy matrix
\begin{align}
\label{first-mon}
T^{(2)}_{\beta}(\m|\lambda_{\ell},\dots,\lambda_1)
=
D^{(1)}_{\beta}(\m)
R^{(2)}_{\beta\alpha_{\ell}}(\m,\lambda_{\ell})
\dots
R^{(2)}_{\beta\alpha_1}(\m,\lambda_1)
=
\left(
\begin{array}{cc}
A^{(2)}(\m|\lambda_{\ell},\dots,\lambda_1)
&
C^{(2)}(\m|\lambda_{\ell},\dots,\lambda_1)
\\
B^{(2)}(\m|\lambda_{\ell},\dots,\lambda_1)
&
D^{(2)}(\m|\lambda_{\ell},\dots,\lambda_1)
\end{array}
\right)_{\beta} 
\end{align}
and each $R^{(2)}_{\beta\alpha_i}(\mu,\lambda_i)$ is given by \r{Rmat-su32}. The reference state in 
(\ref{gen-bethe}) is the tensor product of 
\begin{align*}
|\Uparrow_{\alpha} \rangle 
= 
\bigotimes_{i=1}^{\ell}
\left(
\begin{array}{c} 
1 \\ 0
\end{array}
\right)_{\alpha_i}
\end{align*} 
and the pseudo-vacuum $|0 \rangle$. Notice that $|\b{\l}_{\ell}, \b{\m}_m\rangle \in \mathcal{H}$, since the vector spaces $V_{\alpha_i}$ are purely auxiliary and play no role other than to avoid complicated summation in the Ansatz for the eigenvectors. At this stage, the two sets of variables $\b\l_{\ell}$ and $\b\m_m$ introduced in \r{gen-bethe} are considered to be free. When that is the case, we will refer to elements of $\mathcal{H}$ of the form \r{gen-bethe} as {\it generic Bethe vectors} or {\it off-shell states.} The Ansatz \r{gen-bethe} does not give an eigenvector of \r{transfer} unless the variables $\b\l_{\ell}$ and $\b\m_m$ satisfy the Bethe equations, as we describe in the following subsection. Nevertheless, off-shell states are rather special elements of $\mathcal{H}$ and satisfy some remarkable properties; see in particular equation \r{T-action} in \s{s:review}, which turns out to be essential to our results. 

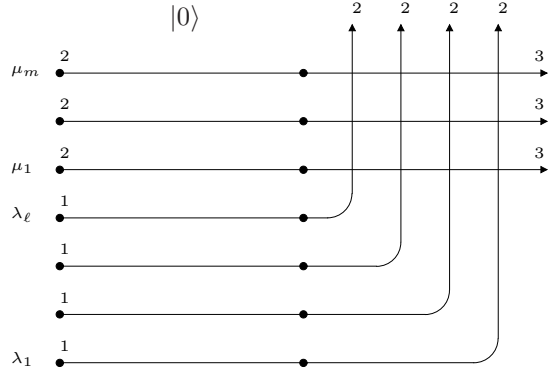
\begin{figure}

\begin{center}
\begin{minipage}{4.3in}

\setlength{\unitlength}{0.00032cm}
\begin{picture}(40000,10000)(14000,10000)


\put(26500,24000){$|0\rangle$}

\blacken\path(41750,18125)(41750,17875)(42000,18000)(41750,18125)
\blacken\path(41750,20125)(41750,19875)(42000,20000)(41750,20125)
\blacken\path(41750,22125)(41750,21875)(42000,22000)(41750,22125)


\path(22000,22000)(32000,22000)
\put(22000,22000){\circle*{300}}
\put(32000,22000){\circle*{300}}
\put(20000,22000){\tiny $\m_m$}
\put(22000,22500){\tiny 2}
\put(41500,22500){\tiny 3}

\path(22000,20000)(32000,20000)
\put(22000,20000){\circle*{300}}
\put(32000,20000){\circle*{300}}
\put(22000,20500){\tiny 2}
\put(41500,20500){\tiny 3}

\path(22000,18000)(32000,18000)
\put(22000,18000){\circle*{300}}
\put(32000,18000){\circle*{300}}
\put(20000,18000){\tiny $\m_1$}
\put(22000,18500){\tiny 2}
\put(41500,18500){\tiny 3}

\path(22000,16000)(32000,16000)
\put(22000,16000){\circle*{300}}
\put(32000,16000){\circle*{300}}
\put(20000,16000){\tiny $\l_{\ell}$}
\put(22000,16500){\tiny 1}

\path(22000,14000)(32000,14000)
\put(22000,14000){\circle*{300}}
\put(32000,14000){\circle*{300}}
\put(22000,14500){\tiny 1}

\path(22000,12000)(32000,12000)
\put(22000,12000){\circle*{300}}
\put(32000,12000){\circle*{300}}
\put(22000,12500){\tiny 1}
 
\path(22000,10000)(32000,10000)
\put(22000,10000){\circle*{300}}
\put(32000,10000){\circle*{300}}
\put(20000,10000){\tiny $\l_{1}$}
\put(22000,10500){\tiny 1}


\path(32000,22000)(42000,22000)
\path(32000,20000)(42000,20000)
\path(32000,18000)(42000,18000)

\path(32000,16000)(33000,16000)
\put(33000,17000){\arc{2000}{0}{1.5708}}
\path(34000,17000)(34000,24000)
\put(34000,24500){\tiny 2}

\path(32000,14000)(35000,14000)
\put(35000,15000){\arc{2000}{0}{1.5708}}
\path(36000,15000)(36000,24000)
\put(36000,24500){\tiny 2}

\path(32000,12000)(37000,12000)
\put(37000,13000){\arc{2000}{0}{1.5708}}
\path(38000,13000)(38000,24000)
\put(38000,24500){\tiny 2}

\path(32000,10000)(39000,10000)
\put(39000,11000){\arc{2000}{0}{1.5708}}
\path(40000,11000)(40000,24000)
\put(40000,24500){\tiny 2}

\blacken\path(34125,23750)(33875,23750)(34000,24000)(34125,23750)
\blacken\path(36125,23750)(35875,23750)(36000,24000)(36125,23750)
\blacken\path(38125,23750)(37875,23750)(38000,24000)(38125,23750)
\blacken\path(40125,23750)(39875,23750)(40000,24000)(40125,23750)

\end{picture}

\end{minipage}
\end{center}

\caption{Graphical representation of the Bethe vector $|\b\l_{\ell}, \b\m_m\rangle$. The horizontal line segments delineated by $\gfrac{i}{\bullet}$ and $\gfrac{j}{\bullet}$ represent elements of the monodromy matrix, $T_{ij}$. All internal bonds are summed over all possible values $1,2,3$. Left-to-right multiplication in \r{gen-bethe} corresponds to bottom-to-top ordering in the diagram, hence the pseudo-vacuum $|0\rangle$ is indicated at the top.}

\label{fig-bv1}

\end{figure}

\begin{figure}

\begin{center}
\begin{minipage}{4.3in}

\setlength{\unitlength}{0.00032cm}
\begin{picture}(40000,14000)(-2000,8000)


\put(17500,7500){$\langle 0|$}

\blacken\path(3750,10125)(3750,9875)(4000,10000)(3750,10125)
\blacken\path(3750,12125)(3750,11875)(4000,12000)(3750,12125)
\blacken\path(3750,14125)(3750,13875)(4000,14000)(3750,14125)


\path(13000,22000)(23000,22000)
\put(13000,22000){\circle*{300}}
\put(23000,22000){\circle*{300}}
\put(22500,22500){\tiny 1}
\put(23500,22000){\tiny $\l_{\ell}$}

\path(13000,20000)(23000,20000)
\put(13000,20000){\circle*{300}}
\put(23000,20000){\circle*{300}}
\put(22500,20500){\tiny 1}

\path(13000,18000)(23000,18000)
\put(13000,18000){\circle*{300}}
\put(23000,18000){\circle*{300}}
\put(22500,18500){\tiny 1}

\path(13000,16000)(23000,16000)
\put(13000,16000){\circle*{300}}
\put(23000,16000){\circle*{300}}
\put(22500,16500){\tiny 1}
\put(23500,16000){\tiny $\l_1$}

\path(13000,14000)(23000,14000)
\put(13000,14000){\circle*{300}}
\put(23000,14000){\circle*{300}}
\put(3000,14500){\tiny 3}
\put(22500,14500){\tiny 2}
\put(23500,14000){\tiny $\m_m$}

\path(13000,12000)(23000,12000)
\put(13000,12000){\circle*{300}}
\put(23000,12000){\circle*{300}}
\put(3000,12500){\tiny 3}
\put(22500,12500){\tiny 2}

\path(13000,10000)(23000,10000)
\put(13000,10000){\circle*{300}}
\put(23000,10000){\circle*{300}}
\put(3000,10500){\tiny 3}
\put(22500,10500){\tiny 2}
\put(23500,10000){\tiny $\m_1$}


\path(6000,22000)(13000,22000)
\put(6000,21000){\arc{2000}{3.142}{4.712}}
\path(5000,21000)(5000,8000)
\put(5000, 7000){\tiny 2}

\path(8000,20000)(13000,20000)
\put(8000,19000){\arc{2000}{3.142}{4.712}}
\path(7000,19000)(7000,8000)
\put(7000, 7000){\tiny 2}

\path(10000,18000)(13000,18000)
\put(10000,17000){\arc{2000}{3.142}{4.712}}
\path(9000,17000)(9000,8000)
\put(9000, 7000){\tiny 2}

\path(12000,16000)(13000,16000)
\put(12000,15000){\arc{2000}{3.142}{4.712}}
\path(11000,15000)(11000,8000)
\put(11000, 7000){\tiny 2}

\path(3000,14000)(13000,14000)
\path(3000,12000)(13000,12000)
\path(3000,10000)(13000,10000)

\blacken\path(5125,8750)(4875,8750)(5000,9000)(5125,8750)
\blacken\path(7125,8750)(6875,8750)(7000,9000)(7125,8750)
\blacken\path(9125,8750)(8875,8750)(9000,9000)(9125,8750)
\blacken\path(11125,8750)(10875,8750)(11000,9000)(11125,8750)

\end{picture}

\end{minipage}
\end{center}

\caption{Graphical representation of the dual Bethe vector $\langle \b\m_m, \b\l_{\ell} |$. As in figure \ref{fig-bv1}, left-to-right multiplication in \r{dual-gen-bethe} corresponds to bottom-to-top ordering in the diagram, hence the dual pseudo-vacuum $\langle 0|$ is indicated at the bottom.}

\label{fig-bv2}

\end{figure}
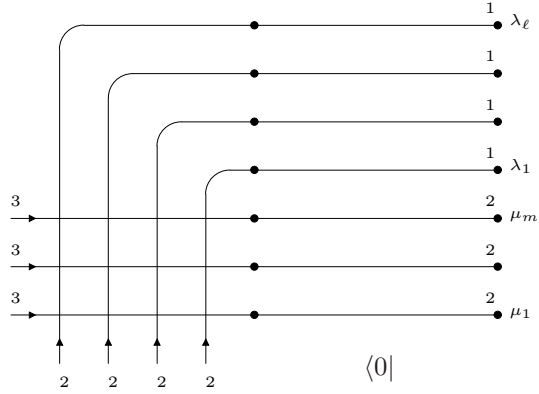

An analogous procedure applies to finding eigenvectors $\langle \Psi| \in \mathcal{H}^{*}$ of \r{transfer}. In that case, the nested Bethe Ansatz leads one to propose that
\begin{align}
\langle \Psi|
=
\langle \b{\m}_m, \b{\l}_{\ell}|
=
\langle \Uparrow_{\alpha}|
\otimes
\langle 0|
B^{(2)}(\mu_1)
\dots
B^{(2)}(\mu_m)
B^{(1)}_{\alpha_1}(\lambda_1)
\dots
B^{(1)}_{\alpha_{\ell}}(\lambda_{\ell})
\label{dual-gen-bethe}
\end{align}
where each $B^{(1)}_{\alpha_i}(\lambda_i)$ is a column vector given by (\ref{decomposition}), and $B^{(2)}(\m)$ denotes the $(2,1)$-th component of the $SU(2)$-invariant monodromy matrix
\begin{align*}
T^{(2)}_{\beta}(\lambda_{\ell},\dots,\lambda_1|\m)
=
R^{(2)}_{\beta\alpha_{\ell}}(\m,\lambda_{\ell})
\dots
R^{(2)}_{\beta\alpha_1}(\m,\lambda_1)
D^{(1)}_{\beta}(\m)
=
\left(
\begin{array}{cc}
A^{(2)}(\lambda_{\ell},\dots,\lambda_1|\m)
&
C^{(2)}(\lambda_{\ell},\dots,\lambda_1|\m)
\\
B^{(2)}(\lambda_{\ell},\dots,\lambda_1|\m)
&
D^{(2)}(\lambda_{\ell},\dots,\lambda_1|\m)
\end{array}
\right)_{\beta} 
\end{align*}
We point out that the only difference between this monodromy matrix and that in \r{first-mon} is the order in which $D^{(1)}_{\beta}(\mu)$ and the $R$-matrices $R^{(2)}_{\beta\alpha_i}(\mu,\lambda_i)$ are multiplied. The reference state in (\ref{dual-gen-bethe}) is the tensor product of 
\begin{align*}
\langle \Uparrow_{\alpha}| 
= 
\bigotimes_{i=1}^{\ell}
\left(
\begin{array}{cc} 
1 & 0
\end{array}
\right)_{\alpha_i}
\end{align*} 
and the dual pseudo-vacuum $\langle 0|$. Similarly to above, if the variables $\b\l_{\ell}$ and $\b\m_m$ are free, we refer to \r{dual-gen-bethe} as {\it dual generic Bethe vectors} or {\it dual off-shell states.}

\subsection{Bethe eigenvectors (on-shell states)}

In order for the Ans\"atze \r{gen-bethe} and \r{dual-gen-bethe} to successfully give the eigenvectors of the transfer matrix \r{transfer}, it is necessary to enforce constraints on the variables $\b\l_{\ell}$ and $\b\m_m$. It is possible to show that \r{gen-bethe} and \r{dual-gen-bethe} are eigenvectors of the transfer matrix, and satisfy the equations
\begin{align*}
\mathcal{T}(z) 
|\b{\l}_{\ell}, \b{\m}_m \rangle
=
\Lambda\Big(z \Big| \b{\l}_{\ell}, \b{\m}_m \Big)
|\b{\l}_{\ell}, \b{\m}_m \rangle,
\quad\quad
\langle \b{\m}_m, \b{\l}_{\ell} |
\mathcal{T}(z)
=
\Lambda\Big(z \Big| \b{\l}_{\ell}, \b{\m}_m \Big)
\langle \b{\m}_m, \b{\l}_{\ell} |
\end{align*}
with the (common) eigenvalue
\begin{align}
\label{eigval}
\Lambda\Big(z \Big| \b{\l}_{\ell},\b{\m}_m \Big)
=
a_1(z)
\prod_{i=1}^{\ell} f(\l_i,z)
+
a_2(z)
\prod_{i=1}^{\ell} f(z,\l_i)
\prod_{j=1}^{m} f(\m_j,z)
+
a_3(z)
\prod_{j=1}^{m} f(z,\m_j)
\end{align}
if and only if the variables $\b{\l}_{\ell}$ and $\b{\m}_m$ satisfy the Bethe equations:
\begin{align}
\label{bethe-l}
r_1(\l_i)
=
\frac{a_1(\l_i)}{a_2(\l_i)}
=
- 
\prod_{k=1}^{\ell}
\left(
\frac{\l_k - \l_i -1}{\l_k - \l_i +1}
\right)
\prod_{k=1}^{m}
\left(
\frac{\m_k - \l_i +1}{\m_k - \l_i}
\right)
\\
\label{bethe-m}
r_3(\m_j)
=
\frac{a_3(\m_j)}{a_2(\m_j)}
=
-
\prod_{k=1}^{m}
\left(
\frac{\m_j - \m_k -1}{\m_j - \m_k +1}
\right)
\prod_{k=1}^{\ell}
\left(
\frac{\m_j - \l_k +1}{\m_j - \l_k}
\right)
\end{align}
In this case we refer to \r{gen-bethe} and \r{dual-gen-bethe} as \emph{Bethe eigenvectors} or \emph{on-shell states}, and to the variables $\b{\l}_{\ell}$ and $\b{\m}_m$ as {\it Bethe roots.} For the sake of our ensuing calculations, it is useful to introduce the functions
\begin{align}
\label{beta1}
\beta_1 \Big( \nu \Big| \b{\l}_{\ell},\b{\m}_m \Big)
=
1+r_1(\nu)
\prod_{j=1}^{m}
\left(
\frac{\mu_j-\nu}{\mu_j-\nu+1}
\right)
\prod_{i=1}^{\ell}
\left(
\frac{\lambda_i-\nu+1}{\lambda_i-\nu-1}
\right)
\\
\label{beta3}
\beta_3 \Big( \nu \Big| \b{\l}_{\ell},\b{\m}_m \Big)
=
1+r_3(\nu) 
\prod_{i=1}^{\ell}
\left(
\frac{\nu-\lambda_i}{\nu-\lambda_i+1}
\right)
\prod_{j=1}^{m}
\left(
\frac{\nu-\mu_j+1}{\nu-\mu_j-1}
\right)
\end{align}
in terms of which the Bethe equations are very simply expressed. If $\b{\l}_{\ell}$ and $\b{\m}_m$ are Bethe roots, then the Bethe equations \r{bethe-l} and \r{bethe-m} can be written as
\begin{align*}
\beta_1 \Big( \lambda_i \Big| \b{\l}_{\ell},\b{\m}_m \Big)
=
0, \quad\forall\ 1\leq i\leq \ell,
\quad
\text{and}
\quad
\beta_3 \Big( \mu_j \Big| \b{\l}_{\ell},\b{\m}_m \Big)
=
0, \quad\forall\ 1\leq j\leq m,
\quad
\text{respectively.}
\end{align*}

\subsection{Bethe vectors of $SU(2)$-invariant models as special cases}
\label{ss:su2-cases}

When the set $\b\m_m$ parametrizing the states \r{gen-bethe} and 
\r{dual-gen-bethe} is empty, they simplify to
\begin{align*}
| \b{\l}_{\ell},\emptyset \rangle
=
T_{12}(\lambda_1) \dots T_{12}(\lambda_{\ell}) |0\rangle,
\quad\quad
\langle \emptyset, \b{\l}_{\ell} |
=
\langle 0| T_{21}(\lambda_{1}) \dots T_{21}(\lambda_{\ell})
\end{align*}
Similarly when the set $\b\l_{\ell}$ is empty, we obtain 
\begin{align*}
| \emptyset, \b{\m}_m \rangle
=
T_{23}(\m_1) \dots T_{23}(\m_m) |0\rangle,
\quad\quad
\langle \b{\m}_m, \emptyset |
=
\langle 0| T_{32}(\m_1) \dots T_{32}(\m_m)
\end{align*}
These can be recognized as Bethe vectors of a generalized $SU(2)$-invariant model, meaning that they are used in the algebraic Bethe Ansatz approach \cite{fst,fad,kbi} for diagonalizing the transfer matrices $T_{11}(z)+T_{22}(z)$ and $T_{22}(z)+T_{33}(z)$, respectively. In either case a single set of Bethe equations \r{bethe-l} or \r{bethe-m} trivializes, and the other set survives:
\begin{align*}
r_1(\lambda_i)
=
- 
\prod_{k=1}^{\ell}
\left(
\frac{\l_k - \l_i -1}{\l_k - \l_i +1}
\right),
\quad
\text{when}\ 
m=0,
\quad\quad
r_3(\m_j)
=
-
\prod_{k=1}^{m}
\left(
\frac{\m_j - \m_k -1}{\m_j - \m_k +1}
\right)
\quad
\text{when}\ 
\ell=0.
\end{align*}

\subsection{Definition of the scalar product}
\label{ss:sp-def}

For convenience, let us define renormalized versions of the generic Bethe vectors \r{gen-bethe} and \r{dual-gen-bethe}:
\begin{align*}
\bvec{\b\l_{\ell}}{\b\m_m} 
= 
\frac{| \b\l_{\ell}, \b\m_m \rangle}
{f(\b\m_m,\b\l_{\ell}) a_2(\b\l_{\ell}) a_2(\b\m_m)},
\quad\quad
\dbvec{\b\m_m}{\b\l_{\ell}}
=
\frac{\langle \b\m_m, \b\l_{\ell} |}
{f(\b\m_m,\b\l_{\ell}) a_2(\b\l_{\ell}) a_2(\b\m_m)}
\end{align*}
where we have adopted the notation 
$a_2(\b\l_{\ell}) = \prod_{i=1}^{\ell} a_2(\l_i)$ and
$a_2(\b\m_m) = \prod_{j=1}^{m} a_2(\m_j)$. The scalar product $\mathcal{S}_{\ell,m}$ is a function in four sets of variables; $\blb_{\ell}$ and $\blc_{\ell}$ of cardinality $\ell$, and $\bmb_m$ and $\bmc_m$ of cardinality $m$. It is given by the action of the dual vector 
$\dbvec{\bmb_m}{\blb_{\ell}} \in \mathcal{H}^{*}$ on the vector 
$\bvec{\blc_{\ell}}{\bmc_m} \in \mathcal{H}$:
\begin{align}
\label{on-off}
\mathcal{S}_{\ell,m}(\bmb, \blb | \blc, \bmc)
=
\llangle \bmb_m, \blb_{\ell} \| \blc_{\ell}, \bmc_m \rrangle
\end{align}
For the moment, let us assume that $\dbvec{\bmb_m}{\blb_{\ell}}$ and $\bvec{\blc_{\ell}}{\bmc_m}$ are both 
off-shell states. Using the definitions \r{gen-bethe} and \r{dual-gen-bethe} of the Bethe vectors, the commutation relations of the algebra \r{int-su3} and the defining properties \r{ps-vac3} and \r{ps-vac4} of a generalized $SU(3)$-invariant model, it is possible to show \cite{res} that the off-shell/off-shell scalar product \r{on-off} must be of the form
\begin{align}
\label{generic-sum}
\mathcal{S}_{\ell,m}( \bmb, \blb | \blc, \bmc )
=
\sum
\mathcal{K}\left( 
\begin{array}{c|c||c|c} 
\lci & \lcii & \mci & \mcii \\ \lbi & \lbii & \mbi & \mbii 
\end{array} 
\right)
r_1(\lbi) r_1(\lcii)
r_3(\mbi) r_3(\mcii)
\end{align}
where the sum is over all partitions of the variables into disjoint subsets
\begin{align}
\label{disj1}
&
\blc = \lci \oplus \lcii, \quad
\blb = \lbi \oplus \lbii, \quad
\text{such that}\ 
|\lbi| = |\lci|, \quad 
|\lbii| = |\lcii|
\\
\label{disj2}
&
\bmc = \mci \oplus \mcii, \quad
\bmb = \mbi \oplus \mbii, \quad
\text{such that}\ 
|\mbi| = |\mci|, \quad 
|\mbii| = |\mcii|
\end{align}
and the coefficients $\mathcal{K}$ are functions of $\blb,\blc,\bmb,\bmc$, which depend on the partitioning.
In that sense, $\mathcal{S}_{\ell,m}$ is to be understood not only as a function in $\blb,\blc,\bmb,\bmc$ but also as a linear function in $r_1(\lb_i),r_1(\lc_i)$ for all $1\leq i \leq \ell$ and 
$r_3(\mb_j), r_3(\mc_j)$ for all $1\leq j \leq m$. It is therefore well defined to consider $\mathcal{S}_{\ell,m}$ with these functions set to some value, or modified in some way, since they simply play the role of variables in \r{generic-sum}.  

In almost all of the calculations which follow, we will treat $\dbvec{\bmb_m}{\blb_{\ell}}$ as a dual on-shell state (meaning that the Bethe equations apply to $\blb_{\ell}$ and $\bmb_m$) and $\bvec{\blc_{\ell}}{\bmc_m}$ as an off-shell state. In that situation, we shall refer to \r{on-off} as the {\it on-shell/off-shell $SU(3)$ scalar product.}

\section{On-shell/off-shell scalar products in $SU(2)$ and $SU(3)$-invariant models}
\label{s:results}

\subsection{Determinant expression for on-shell/off-shell $SU(2)$ scalar product}

In $SU(2)$-invariant models, the scalar product between an on-shell and off-shell state was expressed in determinant form by Slavnov in \cite{sla}. We recall this result briefly. Following from the discussion of \s{ss:su2-cases}, two types of on-shell/off-shell $SU(2)$ scalar product appear as particular cases of \r{on-off}:
\begin{align*}
\mathcal{S}_{\ell,0} (\emptyset, \blb | \blc, \emptyset)
&=
\langle 0| \mathbb{T}_{21}(\lb_1) \dots \mathbb{T}_{21}(\lb_{\ell})
\mathbb{T}_{12}(\lc_1) \dots \mathbb{T}_{12}(\lc_{\ell}) |0\rangle,
\quad
r_1(\lb_i)
=
-\prod_{k=1}^{\ell}
\left(
\frac{\lb_k - \lb_i-1}{\lb_k - \lb_i+1}
\right)
\\
\mathcal{S}_{0,m} (\bmb, \emptyset | \emptyset, \bmc)
&=
\langle 0| \mathbb{T}_{32}(\mb_1) \dots \mathbb{T}_{32}(\mb_m)
\mathbb{T}_{23}(\mc_1) \dots \mathbb{T}_{23}(\mc_m) |0\rangle,
\quad
r_3(\mb_j)
=
-\prod_{k=1}^{m}
\left(
\frac{\mb_j - \mb_k-1}{\mb_j - \mb_k+1}
\right) 
\end{align*}
where we define $\mathbb{T}_{ij}(z) \equiv T_{ij}(z)/a_2(z)$ for all $i,j$. Given that the dual states in these scalar products are on-shell, as indicated by the Bethe equations written alongside, we may replace all instances of $r_1(\lb_i)$ and $r_3(\mb_j)$ in $\mathcal{S}_{\ell,0}$ and $\mathcal{S}_{0,m}$ by equivalent rational functions in $\blb$ and $\bmb$, respectively. After doing so, these scalar products may be expressed in determinant form \cite{sla}: 
\begin{align}
\label{slavnov-l}
\mathcal{S}_{\ell,0} (\emptyset, \blb | \blc, \emptyset)
&=
\frac{1}{ \ivan(\blb) \dvan(\blc) }
\det\left(
S^{(1)}_j ( \emptyset, \blb | \lc_i )
\right)_{1 \leq i,j \leq \ell}
\\
\label{slavnov-m}
\mathcal{S}_{0,m} (\bmb, \emptyset | \emptyset, \bmc)
&=
\frac{1}{ \dvan(\bmb) \ivan(\bmc)}
\det\left(
S^{(3)}_j ( \bmb, \emptyset | \mc_i )
\right)_{1 \leq i,j \leq m}
\end{align}
where the matrix entries are given by
\begin{align}
\label{entries-1}
S^{(1)}_j(\emptyset,\blb | \lc_i)
&=
\frac{1}{\lb_j-\lc_i}
\left(
r_1(\lc_i)
\prod_{k\not=j}^{\ell} (\lb_k-\lc_i+1)
-
\prod_{k\not=j}^{\ell} (\lb_k-\lc_i-1)
\right)
\\
\label{entries-3}
S^{(3)}_j(\bmb, \emptyset | \mc_i)
&=
\frac{1}{\mb_j-\mc_i}
\left(
\prod_{k\not=j}^{m} (\mb_k-\mc_i+1)
-
r_3(\mc_i)
\prod_{k\not=j}^{m} (\mb_k-\mc_i-1)
\right)
\end{align}
and where $\ivan,\dvan$ denote Vandermondes with their product ordered differently:
\begin{align*}
\ivan(\b{x}_n) = \prod_{1 \leq i < j \leq n} (x_i-x_j),
\quad\quad
\dvan(\b{x}_n) = \prod_{1 \leq i < j \leq n} (x_j-x_i)
\end{align*}
Equations \r{slavnov-l} and \r{slavnov-m} are central to this work since they form the bases of the recursion relations which we subsequently derive, relating $\mathcal{S}_{\ell,m}$ to $\mathcal{S}_{\ell,m-1}$ and to $\mathcal{S}_{\ell-1,m}$, respectively. As we have written them, the matrix entries \r{entries-1} and \r{entries-3} are special cases of the functions
\begin{align}
\label{entries-mod-1}
S^{(1)}_j(\bmb,\blb | \lc_i)
&=
\frac{1}{\lb_j-\lc_i}
\left(
r_1(\lc_i)
\prod_{k=1}^{m} 
\left( \frac{\mb_k-\lc_i}{\mb_k-\lc_i+1} \right)
\prod_{k\not=j}^{\ell} (\lb_k-\lc_i+1)
-
\prod_{k\not=j}^{\ell} (\lb_k-\lc_i-1)
\right)
\\
\label{entries-mod-3}
S^{(3)}_j(\bmb, \blb | \mc_i)
&=
\frac{1}{\mb_j-\mc_i}
\left(
\prod_{k\not=j}^{m} (\mb_k-\mc_i+1)
-
r_3(\mc_i)
\prod_{k=1}^{\ell}
\left(
\frac{\mc_i-\lb_k}{\mc_i-\lb_k+1}
\right)
\prod_{k\not=j}^{m} (\mb_k-\mc_i-1)
\right)
\end{align}
which we use in the multiple integral formulae \r{mult-int1} and \r{mult-int2} in the following subsections.

\subsection{First multiple integral expression for on-shell/off-shell $SU(3)$ scalar product}

Define the extended Slavnov-type determinant
\begin{multline}
\label{ext-slavnov}
\mathbb{S}^{(1)}
\Big(\bmb_m, \blb_{\ell} \Big| \blc_{\ell} \Big| \b{x}_m, \b{y}_m \Big)
=
\\
\frac{\phantom{\Big|}g(\bmc,\b{y})/g(\b{x},\bmb)\phantom{\Big|}}
{\ivan(\blb) \dvan(\blc \oplus \bmb)}
\det\left(
\begin{array}{ccc|ccc}
S_1^{(1)}(\bmb, \blb | \lc_1 ) 
&
\cdots
&
S_{\ell}^{(1)}(\bmb, \blb | \lc_1 )
&
g(x_1,\lc_1)
&
\cdots 
&
g(x_m,\lc_1)
\\
\vdots & & \vdots & \vdots & & \vdots
\\
S_1^{(1)}(\bmb, \blb | \lc_{\ell} ) 
&
\cdots
&
S_{\ell}^{(1)}(\bmb, \blb |\lc_{\ell} )
&
g(x_1,\lc_{\ell})
&
\cdots 
&
g(x_m,\lc_{\ell})
\\
\phantom{.} & & \phantom{.} & \phantom{.} & & \phantom{.}
\\
S_1^{(1)} (\bmb, \blb |\mb_1 ) 
&
\cdots
&
S_{\ell}^{(1)}(\bmb, \blb | \mb_1 )
&
g(x_1,\mb_1)
&
\cdots 
&
g(x_m,\mb_1)
\\
\vdots & & \vdots & \vdots & & \vdots
\\
S_1^{(1)}(\bmb, \blb | \mb_m ) 
&
\cdots
&
S_{\ell}^{(1)}(\bmb, \blb | \mb_m )
&
g(x_1,\mb_m)
&
\cdots 
&
g(x_m,\mb_m)
\end{array}
\right)
\end{multline}
The scalar product of an on-shell dual state $\dbvec{\bmb_m}{\blb_{\ell}}$ and an off-shell state 
$\bvec{\blc_{\ell}}{\bmc_m}$ is given by the multiple integral formula
\begin{multline}
\label{mult-int1}
\mathcal{S}_{\ell,m}( \bmb, \blb | \blc, \bmc )
=
\oint_{\mathcal{X}_{m}}
\frac{dx_m}{2\pi\i}
\oint_{\mathcal{Y}_{m}}
\frac{dy_m}{2\pi\i}
\cdots
\oint_{\mathcal{X}_1} 
\frac{dx_1}{2\pi\i}
\oint_{\mathcal{Y}_1}
\frac{dy_1}{2\pi\i}
\ 
\dvan(\b{y})
\ 
\mathbb{S}^{(1)}
\Big(\bmb, \blb \Big| \blc \Big| \b{x}, \b{y} \Big)
\\
\times
\prod_{k=1}^{m}
g(x_k,y_k) h(x_k,\b{X}^{k}) h(\b{Y}^{k},y_k) 
\left(
\frac{\beta_3 (y_k | \b{X}^{k},\b{Y}^{k} )}{g(y_k,\mb_k)}
-
\frac{\beta_1 (x_k | \b{X}^{k},\b{Y}^{k} )}{g(x_k,\mb_k)} 
\right)
g(x_k,\bmb_k)
g(\bmb_k,y_k)
\end{multline}
where $\b{X}^{k}$ and $\b{Y}^{k}$ denote the sets
\begin{align}
\label{xy-sets}
\b{X}^k = \blc_{\ell} \oplus \bmb_{k-1} \ominus \b{x}_{k-1},
\quad\quad
\b{Y}^k = \bmc_m \oplus \bmb_{k-1} \ominus \b{y}_{k-1} 
\end{align}
and the integration contours $\mathcal{X}_k$ and $\mathcal{Y}_k$ surround the points\footnote{Because the integration contours $\mathcal{X}_k$ surround poles at $\bmb_k$, at first glance it appears that \r{mult-int1} depends on the functions $r_1(\mb_j)$, which would contradict \r{generic-sum}. Deeper analysis reveals that in fact these functions cannot appear, due to zeros in either $1/g(x_k,\mb_k)$ or $\beta_1(x_k|\b{X}^k,\b{Y}^k)$. A similar statement applies to \r{mult-int2}, and the functions $r_3(\lb_i)$.}
\begin{align}
\label{contours}
\blc_{\ell} \oplus \bmb_k \subset \mathcal{X}_k,
\quad\quad
\bmc_m \subset \mathcal{Y}_k
\end{align}
%
but no other poles in the integrand.

\subsection{Second multiple integral expression for on-shell/off-shell $SU(3)$ scalar product}

Define another extended Slavnov-type determinant
\begin{multline}
\label{ext-slavnov2}
\mathbb{S}^{(3)}
\Big(\bmb_m, \blb_{\ell} \Big| \bmc_m \Big| \b{x}_{\ell}, \b{y}_{\ell} \Big)
=
\\
\frac{\phantom{\Big|} g(\b{x},\blc) / g(\blb,\b{y}) \phantom{\Big|}}
{\dvan(\bmb) \ivan(\bmc \oplus \blb)}
\det\left(
\begin{array}{ccc|ccc}
g(y_1,\lb_1)
&
\cdots 
&
g(y_{\ell},\lb_1)
&
S_1^{(3)}(\bmb, \blb | \lb_1 ) 
&
\cdots
&
S_{m}^{(3)}(\bmb, \blb | \lb_1 )
\\
\vdots & & \vdots & \vdots & & \vdots
\\
g(y_1,\lb_{\ell})
&
\cdots 
&
g(y_{\ell},\lb_{\ell})
&
S_1^{(3)}(\bmb, \blb | \lb_{\ell} ) 
&
\cdots
&
S_{m}^{(3)}(\bmb, \blb | \lb_{\ell} )
\\
\phantom{.} & & \phantom{.} & \phantom{.} & & \phantom{.}
\\
g(y_1,\mc_1)
&
\cdots 
&
g(y_{\ell},\mc_1)
&
S_1^{(3)}(\bmb, \blb | \mc_1 ) 
&
\cdots
&
S_{m}^{(3)}(\bmb, \blb | \mc_1 )
\\
\vdots & & \vdots & \vdots & & \vdots
\\
g(y_1,\mc_m)
&
\cdots 
&
g(y_{\ell},\mc_m)
&
S_1^{(3)}(\bmb, \blb | \mc_m ) 
&
\cdots
&
S_{m}^{(3)}(\bmb, \blb | \mc_m )
\end{array}
\right)
\end{multline}
in terms of which we have a second multiple integral expression for the on-shell/off-shell scalar product: 
\begin{multline}
\label{mult-int2}
\mathcal{S}_{\ell,m}( \bmb, \blb | \blc, \bmc )
=
\oint_{\mathcal{X}_{\ell}}
\frac{dx_{\ell}}{2\pi\i}
\oint_{\mathcal{Y}_{\ell}}
\frac{dy_{\ell}}{2\pi\i}
\cdots
\oint_{\mathcal{X}_1} 
\frac{dx_1}{2\pi\i}
\oint_{\mathcal{Y}_1}
\frac{dy_1}{2\pi\i}
\ 
\ivan(\b{x})
\ 
\mathbb{S}^{(3)}
\Big(\bmb, \blb \Big| \bmc \Big| \b{x}, \b{y} \Big)
\\
\times
\prod_{k=1}^{\ell}
g(x_k,y_k) h(x_k,\b{X}^{k}) h(\b{Y}^{k},y_k) 
\left(
\frac{\beta_3 (y_k | \b{X}^{k},\b{Y}^{k} )}{g(y_k,\lb_k)}
-
\frac{\beta_1 (x_k | \b{X}^{k},\b{Y}^{k} )}{g(x_k,\lb_k)} 
\right)
g(x_k,\blb_k)
g(\blb_k,y_k)
\end{multline}
where $\b{X}^{k}$ and $\b{Y}^{k}$ denote the sets
\begin{align}
\label{xy-sets2}
\b{X}^k = \blc_{\ell} \oplus \blb_{k-1} \ominus \b{x}_{k-1},
\quad\quad
\b{Y}^k = \bmc_m \oplus \blb_{k-1} \ominus \b{y}_{k-1} 
\end{align}
and the integration contours $\mathcal{X}_k$ and $\mathcal{Y}_k$ surround the points
\begin{align}
\label{contours2}
\blc_{\ell} \subset \mathcal{X}_k,
\quad\quad
\bmc_m \oplus \blb_k \subset \mathcal{Y}_k
\end{align}
%
but no other poles in the integrand.

\subsection{On-shell/off-shell $SU(2)$ scalar products as particular cases}
\label{ss:su2-particular}

It is clear that by fixing the cardinality $m=0$ in \r{mult-int1} and $\ell=0$ in \r{mult-int2} we recover the determinant expressions \r{slavnov-l} and \r{slavnov-m} respectively, for an on-shell/off-shell $SU(2)$ scalar product, as should be the case. On the other hand we can choose $\ell=0$ in \r{mult-int1}, which causes the integration over the contours 
$\mathcal{X}_k$ to trivialize, while the integration over the contours $\mathcal{Y}_k$ remains non-trivial. In this case, we find that
\begin{align}
\label{mult-int2-su2}
\mathcal{S}_{0,m}(\bmb, \emptyset | \emptyset, \bmc)
&=
\oint_{\mathcal{Y}_{m}}
\frac{dy_m}{2\pi\i}
\cdots
\oint_{\mathcal{Y}_1}
\frac{dy_1}{2\pi\i}
\ 
\ivan(\b{y}) (-)^m
\prod_{k=1}^{m}
h(\b{Y}^{k},y_k) 
\beta_3 (y_k | \emptyset,\b{Y}^{k} ) 
g(\bmb_k,y_k)
g(\bmc_m,y_k)
\\
\nonumber
&=
\oint_{\mathcal{Y}_{m}}
\frac{dy_m}{2\pi\i}
\cdots
\oint_{\mathcal{Y}_1}
\frac{dy_1}{2\pi\i}
\ 
\prod_{k=1}^{m}
g(y_k,\mb_k)
f(\b{Y}^{k},y_k) 
\beta_3 (y_k | \emptyset,\b{Y}^{k} ) 
\end{align}
where $\b{Y}^k = \bmc_m \oplus \bmb_{k-1} \ominus \b{y}_{k-1}$ for all $1\leq k \leq m$, as usual, and the final line follows from the fact that 
\begin{align*}
h(\b{Y}^k,y_k) 
=
\frac{h(\bmc_m,y_k) h(\bmb_{k-1},y_k)}
{h(\b{y}_{k-1},y_k)}
\end{align*}
In this way we obtain a multiple integral expression for the on-shell/off-shell $SU(2)$ scalar product, which to the best of our knowledge is new, although a similar formula was also found in \cite{gal1}. Once we succeed in proving \r{mult-int1} in \s{s:recursion}, we will have also shown that \r{slavnov-m} and \r{mult-int2-su2} are equal as rational functions in the variables $\bmb$ and $\bmc$, with each $r_3(\mc_i)$ viewed as a constant.

Similarly, by setting $m=0$ in \r{mult-int2} one obtains
\begin{align}
\label{mult-int1-su2}
\mathcal{S}_{\ell,0}( \emptyset, \blb | \blc, \emptyset)
&=
\oint_{\mathcal{X}_{\ell}}
\frac{dx_{\ell}}{2\pi\i}
\cdots
\oint_{\mathcal{X}_1}
\frac{dx_1}{2\pi\i}
\ 
\dvan(\b{x})
(-)^{\ell}
\prod_{k=1}^{\ell}
h(x_k,\b{X}^{k}) 
\beta_1 (x_k | \b{X}^k, \emptyset ) 
g(x_k,\blb_k)
g(x_k,\blc_{\ell})
\\
\nonumber
&=
\oint_{\mathcal{X}_{\ell}}
\frac{dx_{\ell}}{2\pi\i}
\cdots
\oint_{\mathcal{X}_1}
\frac{dx_1}{2\pi\i}
\ 
\prod_{k=1}^{\ell}
g(\lb_k,x_k)
f(x_k,\b{X}^{k}) 
\beta_1 (x_k | \b{X}^k, \emptyset) 
\end{align}
where $\b{X}^k = \blc_{\ell} \oplus \blb_{k-1} \ominus \b{x}_{k-1}$ for all $1\leq k \leq \ell$, as before. Once again, we remark that by proving \r{mult-int2} one finds as a corollary that \r{slavnov-l} and \r{mult-int1-su2} must be equal as rational functions in the variables $\blb$ and $\blc$, with each $r_1(\lc_i)$ viewed as a constant.

\section{Explicit formulae involving off-shell $SU(3)$ Bethe vectors}
\label{s:review}

In this section we review some results from the paper \cite{bprs3}, which are fundamental to the derivation of our recursion relations for the $SU(3)$ scalar product. Since the proof of these results is quite technical and would require a long digression, we will not present it here. The reader is referred to the original source \cite{bprs3} for more details. 

\subsection{Action of diagonal monodromy matrix elements on generic Bethe vectors}

\begin{myTheorem} The diagonal elements of the monodromy matrix, 
$T_{11}(z)$, $T_{22}(z)$, $T_{33}(z)$, have the following action on Bethe vectors:
\begin{align}
\label{action1}
\mathbb{T}_{11}(z)
\bvec{\b\l}{\b\m}
=
r_1(z) f(\b\l,z) 
\bvec{\b\l}{\b\m}
+
f(\b\m,z) 
\sum_{i=1}^{\ell} 
r_1(\l_i) f^{-1}(\b\m,\l_i) g(z,\l_i) \prod_{k\not=i}^{\ell} f(\l_k,\l_i)
\bvec{\wh\l_i \oplus z}{\b\m}
&
\\
\nonumber
+
\sum_{i=1}^{\ell} \sum_{j=1}^{m}
r_1(\l_i) f^{-1}(\b\m,\l_i) g(z,\m_j) g(\m_j,\l_i)
\prod_{k\not=i}^{\ell} f(\l_k,\l_i)
\prod_{k\not=j}^{m} f(\m_k,\m_j)
\bvec{\wh\l_i \oplus z}{\wh\m_j \oplus z}
& 
\end{align}
\begin{align}
\label{action2}
\mathbb{T}_{22}(z)
\bvec{\b\l}{\b\m}
=
f(z,\b\l) f(\b\m,z) 
\bvec{\b\l}{\b\m}
+
f(\b\m,z) 
\sum_{i=1}^{\ell} 
g(\l_i,z) \prod_{k\not=i}^{\ell} f(\l_i,\l_k)
\bvec{\wh\l_i\oplus z}{\b\m}
&
\\
\nonumber
+
f(z,\b\l) 
\sum_{j=1}^{m} 
g(z,\m_j) \prod_{k\not=j}^{m} f(\m_k,\m_j)
\bvec{\b\l}{\wh\m_j \oplus z}
&
\\
\nonumber
+
\sum_{i=1}^{\ell} \sum_{j=1}^{m}
g(\l_i,z) g(z,\m_j)
\prod_{k\not=i}^{\ell} f(\l_i,\l_k)
\prod_{k\not=j}^{m} f(\m_k,\m_j) 
\bvec{\wh\l_i \oplus z}{\wh\m_j \oplus z}
&
\end{align}
\begin{align}
\label{action3}
\mathbb{T}_{33}(z) 
\bvec{\b\l}{\b\m}
=
r_3(z) f(z,\b\m) 
\bvec{\b\l}{\b\m}
+
f(z,\b\l) 
\sum_{j=1}^{m}
r_3(\m_j) f^{-1}(\m_j,\b\l) 
g(\m_j,z) \prod_{k\not=j}^{m} f(\m_j,\m_k)
\bvec{\b\l}{\wh\m_j \oplus z}
&
\\
\nonumber
+
\sum_{i=1}^{\ell} \sum_{j=1}^{m}
r_3(\m_j) f^{-1}(\m_j,\b\l) 
g(\l_i,z) g(\m_j,\l_i)
\prod_{k\not=i}^{\ell} f(\l_i,\l_k)
\prod_{k\not=j}^{m} f(\m_j,\m_k) 
\bvec{\wh\l_i \oplus z}{\wh\m_j \oplus z}
&
\end{align}
where in all cases $\mathbb{T}_{ii}(z) = T_{ii}(z)/a_2(z)$. These formulae are special cases of those obtained in \cite{bprs3}, in which the action of arbitrarily many elements $\mathbb{T}_{ii}(z_j)$ on a Bethe vector was calculated. 
\end{myTheorem}

\subsection{Action of transfer matrix on generic Bethe vectors}

Summing equations \r{action1}--\r{action3}, we obtain the action of the transfer matrix $\sum_{k=1}^{3} \mathbb{T}_{kk}(z)$ on an off-shell state $\bvec{\b\l}{\b\m}$:
\begin{align}
\label{T-action}
&
\sum_{k=1}^{3} \mathbb{T}_{kk}(z)
\bvec{\b\l}{\b\m}
=
\frac{\Lambda(z| \b\l, \b\m)}{a_2(z)}
\bvec{\b\l}{\b\m}
\\
\nonumber
&+
f(\b\m,z) \sum_{i=1}^{\ell}
g(\l_i,z)
\left(
\prod_{k\not=i}^{\ell}
f(\l_i,\l_k)
-
\frac{r_1(\l_i)}{f(\b\m,\l_i)}
\prod_{k\not=i}^{\ell}
f(\l_k,\l_i)
\right)
\bvec{\wh\l_i \oplus z}{\b\m}
\\
\nonumber
&+
f(z,\b\l) \sum_{j=1}^{m}
g(\m_j,z)
\left(
\frac{r_3(\m_j)}{f(\m_j,\b\l)}
\prod_{k\not=j}^{m}
f(\m_j,\m_k)
-
\prod_{k\not=j}^{m}
f(\m_k,\m_j)
\right)
\bvec{\b\l}{\wh\m_j \oplus z}
\\
\nonumber
&+
\sum_{i=1}^{\ell} \sum_{j=1}^{m}
g(\m_j,z) g(\m_j,\l_i)
\left(
\prod_{k\not=i}^{\ell}
f(\l_i,\l_k)
-
\frac{r_1(\l_i)}{f(\b\m,\l_i)}
\prod_{k\not=i}^{\ell}
f(\l_k,\l_i)
\right)
\prod_{k\not=j}^{m}
f(\m_k,\m_j)
\bvec{\wh\l_i \oplus z}{\wh\m_j \oplus z}
\\
\nonumber
&+
\sum_{i=1}^{\ell} \sum_{j=1}^{m}
g(\l_i,z) g(\m_j,\l_i)
\left(
\frac{r_3(\m_j)}{f(\m_j,\b\l)}
\prod_{k\not=j}^{m}
f(\m_j,\m_k)
-
\prod_{k\not=j}^{m}
f(\m_k,\m_j)
\right)
\prod_{k\not=i}^{\ell}
f(\l_i,\l_k)
\bvec{\wh\l_i \oplus z}{\wh\m_j \oplus z}
\end{align}
where we have used the identity $g(\l_i,z) g(z,\m_j) = -g(z,\m_j) g(\m_j,\l_i)-g(\l_i,z) g(\m_j,\l_i)$ to split the final term in \r{action2} into two terms, which can then be recombined with the final terms of \r{action1} and \r{action3}, respectively. This formula was used in \cite{bprs3} to explicitly recover the nested Bethe equations \r{bethe-l} and \r{bethe-m}, since all coefficients in the sums of \r{T-action} vanish when the Bethe equations are obeyed, and we are left with just the leading term.

\section{Recursion relations for $\mathcal{S}_{\ell,m}(\bmb,\blb | \blc,\bmc )$}
\label{s:recursion}

In this section we derive recursion relations for the scalar product $\mathcal{S}_{\ell,m}$, which specify it entirely in terms of scalar products $\mathcal{S}_{\ell-1,m}$ or $\mathcal{S}_{\ell,m-1}$ \emph{without} specializing any of the variables which appear in $\mathcal{S}_{\ell,m}$. To achieve this, the tools which we need are the formula \r{T-action}, that allows us to compute the action of the transfer matrix on a generic Bethe state, and the recursion relations \r{res1} and \r{res2}, which relate the residues of $\mathcal{S}_{\ell,m}$ at certain poles to modified scalar products $\mathcal{S}_{\ell-1,m}$ and $\mathcal{S}_{\ell,m-1}$.

Once we have obtained our recursion relations, it is a straightforward calculation to prove that the multiple integral formulae \r{mult-int1} and \r{mult-int2} are indeed solutions.   

\subsection{Sum form for $SU(3)$ scalar product}

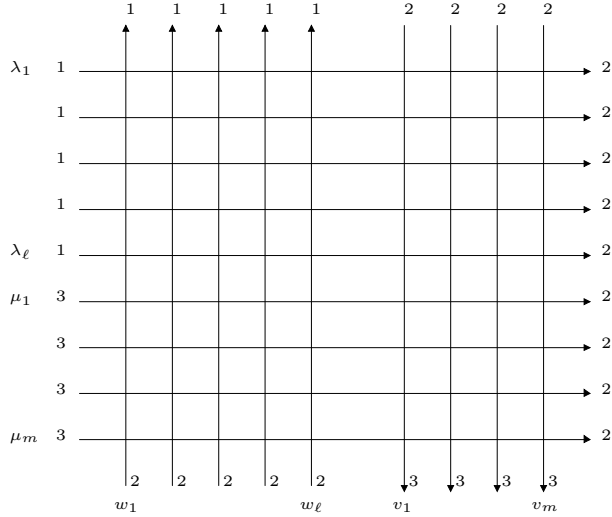
\begin{figure}

\begin{center}
\begin{minipage}{4.3in}

\setlength{\unitlength}{0.0003cm}
\begin{picture}(40000,21000)(8000,-7000)

\blacken\path(35750,12125)(35750,11875)(36000,12000)(35750,12125)
\blacken\path(35750,10125)(35750,9875)(36000,10000)(35750,10125)
\blacken\path(35750,8125)(35750,7875)(36000,8000)(35750,8125)
\blacken\path(35750,6125)(35750,5875)(36000,6000)(35750,6125)
\blacken\path(35750,4125)(35750,3875)(36000,4000)(35750,4125)
\blacken\path(35750,2125)(35750,1875)(36000,2000)(35750,2125)
\blacken\path(35750,0125)(35750,-0125)(36000,0000)(35750,0125)
\blacken\path(35750,-1825)(35750,-2125)(36000,-2000)(35750,-1825)
\blacken\path(35750,-3825)(35750,-4125)(36000,-4000)(35750,-3825)


\put(11000,12000){\tiny $\lambda_1$}
\put(13000,12000){\tiny 1}
\path(14000,12000)(36000,12000)
\put(36500,12000){\tiny 2} 

\put(13000,10000){\tiny 1}
\path(14000,10000)(36000,10000) 
\put(36500,10000){\tiny 2}

\put(13000,8000){\tiny 1}
\path(14000,8000)(36000,8000)
\put(36500,8000){\tiny 2} 

\put(13000,6000){\tiny 1}
\path(14000,6000)(36000,6000)
\put(36500,6000){\tiny 2} 

\put(11000,4000){\tiny $\lambda_{\ell}$}
\put(13000,4000){\tiny 1}
\path(14000,4000)(36000,4000) 
\put(36500,4000){\tiny 2}


\put(11000,2000){\tiny $\mu_1$}
\put(13000,2000){\tiny 3}
\path(14000,2000)(36000,2000)
\put(36500,2000){\tiny 2} 

\put(13000,000){\tiny 3}
\path(14000,000)(36000,000)
\put(36500,000){\tiny 2} 

\put(13000,-2000){\tiny 3}
\path(14000,-2000)(36000,-2000)
\put(36500,-2000){\tiny 2} 

\put(11000,-4000){\tiny $\mu_m$}
\put(13000,-4000){\tiny 3}
\path(14000,-4000)(36000,-4000)
\put(36500,-4000){\tiny 2}


\path(16000,-6000)(16000,14000)
\put(15500,-7000){\tiny $w_1$}
\put(16200,-6000){\tiny 2} 
\put(16000,14500){\tiny 1} 

\path(18000,-6000)(18000,14000)
\put(18200,-6000){\tiny 2} 
\put(18000,14500){\tiny 1} 

\path(20000,-6000)(20000,14000)
\put(20200,-6000){\tiny 2} 
\put(20000,14500){\tiny 1} 

\path(22000,-6000)(22000,14000)
\put(22200,-6000){\tiny 2} 
\put(22000,14500){\tiny 1} 

\path(24000,-6000)(24000,14000) 
\put(23500,-7000){\tiny $w_{\ell}$}
\put(24200,-6000){\tiny 2} 
\put(24000,14500){\tiny 1}


\path(28000,-6000)(28000,14000)
\put(27500,-7000){\tiny $v_1$}
\put(28200,-6000){\tiny 3} 
\put(28000,14500){\tiny 2} 

\path(30000,-6000)(30000,14000)
\put(30200,-6000){\tiny 3} 
\put(30000,14500){\tiny 2} 

\path(32000,-6000)(32000,14000)
\put(32200,-6000){\tiny 3} 
\put(32000,14500){\tiny 2}

\path(34000,-6000)(34000,14000)
\put(33500,-7000){\tiny $v_m$}
\put(34200,-6000){\tiny 3} 
\put(34000,14500){\tiny 2}

\blacken\path(16125,13750)(15875,13750)(16000,14000)(16125,13750)
\blacken\path(18125,13750)(17875,13750)(18000,14000)(18125,13750)
\blacken\path(20125,13750)(19875,13750)(20000,14000)(20125,13750)
\blacken\path(22125,13750)(21875,13750)(22000,14000)(22125,13750)
\blacken\path(24125,13750)(23875,13750)(24000,14000)(24125,13750)

\blacken\path(28125,-6000)(27875,-6000)(28000,-6250)(28125,-6000)
\blacken\path(30125,-6000)(29875,-6000)(30000,-6250)(30125,-6000)
\blacken\path(32125,-6000)(31875,-6000)(32000,-6250)(32125,-6000)
\blacken\path(34125,-6000)(33875,-6000)(34000,-6250)(34125,-6000)

\end{picture}

\end{minipage}
\end{center}

\caption{Lattice representation of $Z(\b\lambda,\b\mu|\b{w},\b{v})$. Summation is implied on all internal line segments in the lattice. Each intersection of lattice lines is a vertex of the type in figure \ref{f:vert1}, where it is important to be mindful of the differing orientations of the vertical lines. For a detailed derivation of the sum formula \r{sum-su3}, we refer the reader to \cite{whe}.}
\label{f:resh-pf}

\end{figure}

As we have already mentioned in \s{ss:sp-def}, it is possible to derive an explicit sum formula for the off-shell/off-shell scalar product in $SU(3)$-invariant models, provided one is able to calculate the coefficients $\mathcal{K}$ appearing in \r{generic-sum}. In \cite{res}, Reshetikhin was able to do precisely that, to obtain the following formula: 
\begin{multline}
\label{sum-su3}
f(\bmb,\blb) f(\bmc,\blc) 
\mathcal{S}_{\ell,m} (\bmb, \blb | \blc, \bmc)
=
\sum
Z(\lbii,\mci | \lcii,\mbi)
Z(\lci,\mbii | \lbi,\mcii) 
\\
\times
f(\lci,\lcii) f(\lbii,\lbi) 
f(\mcii,\mci) f(\mbi,\mbii)
f(\mbii,\lbii) f(\mci,\lci)
r_1(\lbi) r_1(\lcii)
r_3(\mbi) r_3(\mcii)
\end{multline}
where the sum is over the same partitioning as \r{disj1} and \r{disj2}. In the original article \cite{res}, the function $Z$ which appears in \r{sum-su3} was expressed as the partition function of the lattice in figure \ref{f:resh-pf}. An explicit formula for $Z$ was found in \cite{whe}: 
\begin{align}
\label{resh-pf}
Z(\b{\lambda},\b{\mu}|\b{w},\b{v})
=
\sum_{
\b{\lambda} = \li \oplus \lii,\ 
\b{\mu} = \mi \oplus \mii
}
f(\mi,\mii)
f(\lii,\li)
f(\mi,\li)
K(\lii|\mii)
K(\li \oplus \mii|\b{w})
K(\b{v}| \mi \oplus \lii)
\end{align}
where the sum is taken over all partitionings $\b\l = \li \oplus \lii$ and $\b\m = \mi \oplus \mii$, such that $|\lii| = |\mii|$, and where $K(\b{x} | \b{y})$ is the domain wall partition function of the rational six-vertex model \cite{kor}, given by the Izergin determinant formula \cite{ize}:
\begin{align*}
K( \b{x}_{\ell} | \b{y}_{\ell} )
=
\frac{\prod_{i,j=1}^{\ell} (x_i-y_j+1)}
{\ivan(\b{x}) \dvan(\b{y})}
\det\left(
\frac{1}{(x_i-y_j+1)(x_i-y_j)}
\right)_{1\leq i,j \leq \ell}
\end{align*}
%

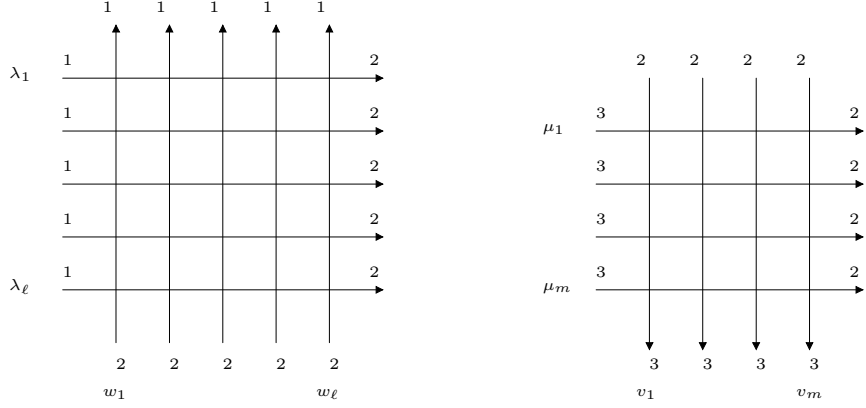
\begin{figure}

\begin{center}
\begin{minipage}{4.3in}

\setlength{\unitlength}{0.00035cm}
\begin{picture}(40000,13500)(13000,8500)


\blacken\path(16125,21750)(15875,21750)(16000,22000)(16125,21750)
\blacken\path(18125,21750)(17875,21750)(18000,22000)(18125,21750)
\blacken\path(20125,21750)(19875,21750)(20000,22000)(20125,21750)
\blacken\path(22125,21750)(21875,21750)(22000,22000)(22125,21750)
\blacken\path(24125,21750)(23875,21750)(24000,22000)(24125,21750)

\blacken\path(25750,12125)(25750,11875)(26000,12000)(25750,12125)
\blacken\path(25750,14125)(25750,13875)(26000,14000)(25750,14125)
\blacken\path(25750,16125)(25750,15875)(26000,16000)(25750,16125)
\blacken\path(25750,18125)(25750,17875)(26000,18000)(25750,18125)
\blacken\path(25750,20125)(25750,19875)(26000,20000)(25750,20125)


\path(14000,20000)(26000,20000) \put(12000,20000){\tiny $\lambda_1$} 
\put(14000,20500){\tiny 1}  
\put(25500,20500){\tiny 2}

\path(14000,18000)(26000,18000)
\put(14000,18500){\tiny 1}  
\put(25500,18500){\tiny 2}

\path(14000,16000)(26000,16000)
\put(14000,16500){\tiny 1}  
\put(25500,16500){\tiny 2}

\path(14000,14000)(26000,14000)
\put(14000,14500){\tiny 1}  
\put(25500,14500){\tiny 2}

\path(14000,12000)(26000,12000) \put(12000,12000){\tiny $\lambda_{\ell}$}
\put(14000,12500){\tiny 1} 
\put(25500,12500){\tiny 2}


\path(16000,10000)(16000,22000) \put(15500,8000){\tiny $w_1$}
\put(16000,9000){\tiny 2} \put(15500,22500){\tiny 1}

\path(18000,10000)(18000,22000)
\put(18000,9000){\tiny 2} \put(17500,22500){\tiny 1}

\path(20000,10000)(20000,22000)
\put(20000,9000){\tiny 2} \put(19500,22500){\tiny 1}

\path(22000,10000)(22000,22000)
\put(22000,9000){\tiny 2} \put(21500,22500){\tiny 1}

\path(24000,10000)(24000,22000) \put(23500,8000){\tiny $w_{\ell}$}
\put(24000,9000){\tiny 2} \put(23500,22500){\tiny 1}


\blacken\path(36125,10000)(35875,10000)(36000,9750)(36125,10000)
\blacken\path(38125,10000)(37875,10000)(38000,9750)(38125,10000)
\blacken\path(40125,10000)(39875,10000)(40000,9750)(40125,10000)
\blacken\path(42125,10000)(41875,10000)(42000,9750)(42125,10000)

\blacken\path(43750,12125)(43750,11875)(44000,12000)(43750,12125)
\blacken\path(43750,14125)(43750,13875)(44000,14000)(43750,14125)
\blacken\path(43750,16125)(43750,15875)(44000,16000)(43750,16125)
\blacken\path(43750,18125)(43750,17875)(44000,18000)(43750,18125)


\path(34000,18000)(44000,18000) \put(32000,18000){\tiny $\m_1$}
\put(34000,18500){\tiny 3}  
\put(43500,18500){\tiny 2}

\path(34000,16000)(44000,16000)
\put(34000,16500){\tiny 3}  
\put(43500,16500){\tiny 2}

\path(34000,14000)(44000,14000)
\put(34000,14500){\tiny 3}  
\put(43500,14500){\tiny 2}

\path(34000,12000)(44000,12000) \put(32000,12000){\tiny $\m_m$}
\put(34000,12500){\tiny 3} 
\put(43500,12500){\tiny 2}


\path(36000,10000)(36000,20000) \put(35500,8000){\tiny $v_1$}
\put(36000,9000){\tiny 3} \put(35500,20500){\tiny 2}

\path(38000,10000)(38000,20000)
\put(38000,9000){\tiny 3} \put(37500,20500){\tiny 2}

\path(40000,10000)(40000,20000)
\put(40000,9000){\tiny 3} \put(39500,20500){\tiny 2}

\path(42000,10000)(42000,20000) \put(41500,8000){\tiny $v_m$}
\put(42000,9000){\tiny 3} \put(41500,20500){\tiny 2}

\end{picture}

\end{minipage}
\end{center}

\caption{On the left, the lattice representation of $Z(\b\l,\emptyset |\b{w},\emptyset)$; on the right, that of $Z(\emptyset,\b\m |\emptyset,\b{v})$. In both cases, we obtain a domain wall partition function of the six-vertex model. On the left, we have $K(\b\l|\b{w})$; on the right, $K(\b{v}|\b\m)$.}
\label{f:dwpf}

\end{figure}

A number of alternative expressions for $Z$, of an analogous nature to \r{resh-pf}, were also found in \cite{bprs1}. Observe that the domain wall partition function is recovered as a special case of 
$Z(\b\l, \b\m | \b{w}, \b{v})$, since we have both 
$Z(\b\l, \emptyset | \b{w}, \emptyset) = K(\b\l | \b{w})$ and 
$Z(\emptyset, \b\m | \emptyset, \b{v}) = K(\b{v} | \b\m)$. 
These relations are easily deduced either from figure \ref{f:dwpf}, or as specializations of \r{resh-pf}. In view of these relations, we recover as special cases of \r{sum-su3} the formulae
\begin{align}
\label{su2-sp-l}
\mathcal{S}_{\ell,0} (\emptyset, \blb | \blc, \emptyset)
&=
\sum
K(\lbii | \lcii) K(\lci | \lbi)
f(\lci,\lcii) f(\lbii,\lbi) r_1(\lbi) r_1(\lcii)
\\
\label{su2-sp-m}
\mathcal{S}_{0,m} (\bmb, \emptyset | \emptyset, \bmc)
&=
\sum
K(\mbi | \mci) K(\mcii | \mbii)
f(\mcii,\mci) f(\mbi,\mbii) r_3(\mbi) r_3(\mcii)
\end{align}
which are two copies of the sum formula for the off-shell/off-shell $SU(2)$ scalar product due to Korepin and Izergin \cite{kor,ik}.

\subsection{Residues of the poles in $Z$}

The function $Z(\b\l_{\ell}, \b\m_m | \b{w}_{\ell}, \b{v}_m)$ possesses simple poles at the points $\l_i = w_j$ and $\m_i = v_j$. Since $Z$ is symmetric separately in each of its sets of variables, it suffices to consider just the residues of the poles at $\l_{\ell} = w_{\ell}$ and 
$\m_m = v_m$:
\begin{align}
\label{res-resh1}
\lim_{\l_{\ell},w_{\ell} \rightarrow x}
\Big\{ (\l_{\ell} - w_{\ell}) 
Z(\b\l_{\ell}, \b\m_m | \b{w}_{\ell}, \b{v}_m) \Big\}
&=
f(\b\l_{\ell-1},x) f(\b\m_m,x) f(x,\b{w}_{\ell-1})  
Z(\b\l_{\ell-1}, \b\m_m | \b{w}_{\ell-1}, \b{v}_m)
\\
\label{res-resh2}
\lim_{\m_m,v_m \rightarrow x}
\Big\{ (v_m - \m_m) 
Z(\b\l_{\ell}, \b\m_m | \b{w}_{\ell}, \b{v}_m) \Big\}
&=
f(x,\b\l_{\ell}) f(x,\b\m_{m-1}) f(\b{v}_{m-1},x)  
Z(\b\l_{\ell}, \b\m_{m-1} | \b{w}_{\ell}, \b{v}_{m-1})
\end{align}
These recursion relations, which first appeared in \cite{res}, can be derived from the graphical definition of $Z$. Alternatively, it is possible to show that the explicit formula \r{resh-pf} is a solution of \r{res-resh1} and \r{res-resh2}.

\subsection{Residues of the poles in $\mathcal{S}_{\ell,m}$}

Due to the poles of the functions $Z$ participating in \r{sum-su3}, the off-shell/off-shell scalar product $\mathcal{S}_{\ell,m}$ also possesses simple poles at the points 
$\lc_i = \lb_j$ and $\mc_i = \mb_j$. Starting from \r{sum-su3} and using the recursive behaviour of $Z$ at its poles \r{res-resh1} and \r{res-resh2}, one can derive similar relations for the residues of $\mathcal{S}_{\ell,m}$:
\begin{multline}
\label{res1}
\lim_{\lc_{\ell},\lb_{\ell} \rightarrow \lambda}
\Big\{ (\lc_{\ell}-\lb_{\ell})
\mathcal{S}_{\ell,m}( \bmb_m, \blb_{\ell} | \blc_{\ell}, \bmc_m )
\Big\}
\\
=
\Big( r_1(\lb_{\ell}) - r_1(\lc_{\ell}) \Big)
\prod_{i=1}^{\ell-1} f(\lc_i,\lambda) f(\lb_i,\lambda) 
\mathcal{S}_{\ell-1,m}^{\mod{1}{\lambda}} 
\Big( 
\bmb_{m}, \blb_{\ell-1} \Big| \blc_{\ell-1}, \bmc_{m}  
\Big)
\end{multline}
\begin{multline}
\label{res2}
\lim_{\mc_m,\mb_m \rightarrow \mu}
\Big\{ (\mc_m-\mb_m)
\mathcal{S}_{\ell,m}( \bmb_m, \blb_{\ell} | \blc_{\ell}, \bmc_m )
\Big\}
\\
=
\Big( r_3(\mc_m) - r_3(\mb_m) \Big)
\prod_{j=1}^{m-1} f(\mu,\mc_j) f(\mu,\mb_j) 
\mathcal{S}_{\ell,m-1}^{\mod{3}{\mu}} 
\Big( 
\bmb_{m-1}, \blb_{\ell} \Big| \blc_{\ell}, \bmc_{m-1} 
\Big)
\end{multline}
where $\mathcal{S}_{\ell-1,m}^{\mod{1}{\lambda}}$ and $\mathcal{S}_{\ell,m-1}^{\mod{3}{\mu}}$ denote scalar products as given by \r{sum-su3}, but with the following substitution performed on all functions $r_1$ and $r_3$:
\begin{align}
\label{mod1}
r_1(x) 
\mapsto
r_1(x)
\frac{f(\l,x)}{f(x,\l)},
\quad\quad
r_3(x)
\mapsto
\frac{r_3(x)}{f(x,\lambda)},
\quad\quad
\text{in}\quad 
\mathcal{S}_{\ell-1,m}^{\mod{1}{\lambda}}
\\
\label{mod3}
r_3(y)
\mapsto
r_3(y)
\frac{f(y,\mu)}{f(\mu,y)},
\quad\quad
r_1(y) 
\mapsto
\frac{r_1(y)}{f(\mu,y)},
\quad\quad
\text{in}\quad 
\mathcal{S}_{\ell,m-1}^{\mod{3}{\mu}}
\end{align}
Once again, the formulae \r{res1} and \r{res2} are due to Reshetikhin in \cite{res}. In addition to these facts, subsequently we will also make use of the fact that $\mathcal{S}_{\ell,m}$ {\it does not} have poles at $\lc_i = \mb_j$ or $\mc_i = \lb_j$:
\begin{align}
\label{triv-res}
\lim_{\lc_{\ell} \rightarrow \mb_m}
\Big\{ (\lc_{\ell}-\mb_m)
\mathcal{S}_{\ell,m}( \bmb_m, \blb_{\ell} | \blc_{\ell}, \bmc_m )
\Big\}
=
0,
\quad\quad
\lim_{\mc_m \rightarrow \lb_{\ell}}
\Big\{ (\mc_m-\lb_{\ell})
\mathcal{S}_{\ell,m}( \bmb_m, \blb_{\ell} | \blc_{\ell}, \bmc_m )
\Big\}
=
0
\end{align}
The analyticity of $\mathcal{S}_{\ell,m}$ at these points follows immediately from the fact that $Z(\b\l,\b\m | \b{w}, \b{v})$ does not have poles at $\l_i = \m_j$ or $w_i = v_j$.

\subsection{Expectation value of the transfer matrix and recursion relations}

Let us consider the quantity $\mathcal{S}_{\ell,m}(z)$, defined as the expectation value of the transfer matrix between a dual Bethe eigenvector $\dbvec{\bmb_m}{\blb_{\ell}}$ and a generic Bethe vector 
$\bvec{\blc_{\ell}}{\bmc_m}$:
\begin{align*}
\mathcal{S}_{\ell,m}(z)
\dfn
\sum_{k=1}^{3}
\dbvec{\bmb_m}{\blb_{\ell}} 
\mathbb{T}_{kk}(z) 
\bvec{\blc_{\ell}}{\bmc_m}
\end{align*}
This quantity has simple poles at the points $z=\lb_i$ and $z=\mb_j$, $1\leq i \leq \ell$ and $1\leq j \leq m$, and we consider the residues of the function $\mathcal{S}_{\ell,m}(z)$ at these poles. For definiteness, we restrict our attention to the points $z=\lb_{\ell}$ and $z=\mb_m$.

The residues in question can be calculated by two different approaches. The first, and simplest approach is to act with the transfer matrix on the dual Bethe eigenvector $\dbvec{\bmb_m}{\blb_{\ell}}$, which gives
\begin{align*}
\mathcal{S}_{\ell,m}(z)
=
\frac{\Lambda(z | \blb,\bmb)}{a_2(z)}
\llangle \bmb_m,\blb_{\ell} \| \blc_{\ell},\bmc_m \rrangle
\end{align*}
where the eigenvalue $\Lambda(z | \blb,\bmb)$ is given by \r{eigval}. It is then easy to calculate the residues as follows:
\begin{align}
\label{act-left-l}
\lim_{z \rightarrow \lb_{\ell}}
\Big\{
(z-\lb_{\ell}) \mathcal{S}_{\ell,m}(z)
\Big\}
&=
\left(
\prod_{i=1}^{\ell-1}
f(\lb_{\ell},\lb_i)
\prod_{j=1}^{m}
f(\mb_j,\lb_{\ell})
-
r_1(z)
\prod_{i=1}^{\ell-1}
f(\lb_i,\lb_{\ell})
\right)
\mathcal{S}_{\ell,m}( \bmb,\blb |\blc,\bmc )
\\
\label{act-left-m}
\lim_{z \rightarrow \mb_{m}}
\Big\{
(z-\mb_{m}) \mathcal{S}_{\ell,m}(z)
\Big\}
&=
\left(
r_3(z)
\prod_{j=1}^{m-1}
f(\mb_m,\mb_j)
-
\prod_{j=1}^{m-1}
f(\mb_j,\mb_m)
\prod_{i=1}^{\ell}
f(\mb_m,\lb_i)
\right)
\mathcal{S}_{\ell,m}( \bmb,\blb |\blc,\bmc )
\end{align}
In both cases, the residue is proportional to the on-shell/off-shell scalar product, which is ultimately what allows us to obtain a recursion relation for this scalar product.

The second approach is to act with the transfer matrix on the generic Bethe vector 
$\bvec{\blc_{\ell}}{\bmc_m}$, using the formula \r{T-action}, before taking the limit required to compute the residue. Many terms are accumulated in this process, each of them being a scalar product between the dual Bethe eigenvector 
$\dbvec{\bmb_m}{\blb_{\ell}}$ and a generic Bethe vector $\bvec{\b{x}_{\ell}}{\b{y}_m}$, where  
\begin{align}
\label{x-sets}
\b{x} = \blc, 
\quad\quad\text{or}\quad\quad
\b{x} = \wlc_{i} \oplus z
\\
\label{y-sets}
\b{y} = \bmc, 
\quad\quad\text{or}\quad\quad
\b{y} = \wmc_{j} \oplus z 
\end{align}
but some of these scalar products make no contribution to the residue being computed, since they do not have a pole at the point under consideration; see equation \r{triv-res}. Indeed, all contribution to the residue of the pole at $z=\lb_{\ell}$ comes from scalar products of the form $\llangle \bmb,\blb \| \wlc_i \oplus z,\b{y} \rrangle$, where $\b{y}$ is either of the sets \r{y-sets}. Similarly, all contribution to the residue of the pole at $z=\mb_m$ comes from scalar products of the form $\llangle \bmb,\blb| \b{x},\wmc_j \oplus z \rrangle$, where $\b{x}$ is either of the sets \r{x-sets}. Using the formulae \r{res1} and \r{res2}, we calculate these contributions explicitly:
\begin{align}
\label{contr1}
&
\lim_{z \rightarrow \lb_{\ell}}
\Big\{ (z-\lb_{\ell}) 
\mathcal{S}_{\ell,m} 
\Big( \bmb, \blb | \wlc_i \oplus z, \b{y}  \Big)
\Big\}
\\
\nonumber
&
=
\Big(
r_1(\lb_{\ell}) - r_1(z)
\Big)
\prod_{k\not=i}^{\ell}
f(\lc_k,\lb_{\ell})
\prod_{k=1}^{\ell-1}
f(\lb_k,\lb_{\ell})
\mathcal{S}_{\ell-1,m}^{\mod{1}{\lb_{\ell}}}
\Big(
\bmb, \wlb_{\ell} \Big| \wlc_i, \b{y} 
\Big)
\\
\nonumber
&
=
\left(
f(\bmb,\lb_{\ell})
\prod_{k=1}^{\ell-1}
f(\lb_{\ell},\lb_k)
-
r_1(z)
\prod_{k=1}^{\ell-1}
f(\lb_k,\lb_{\ell})
\right)
\prod_{k\not=i}^{\ell}
f(\lc_k,\lb_{\ell})
\mathcal{S}_{\ell-1,m}^{\mod{1}{\lb_{\ell}}}
\Big(
\bmb, \wlb_{\ell} \Big| \wlc_i, \b{y} 
\Big)
\end{align}

\begin{align}
\label{contr2}
&
\lim_{z \rightarrow \mb_m}
\Big\{ (z-\mb_m) 
\mathcal{S}_{\ell,m} 
\Big( \bmb, \blb | \b{x}, \wmc_j \oplus z  \Big)
\Big\}
\\
\nonumber
&
=
\Big( r_3(z) - r_3(\mb_m) \Big)
\prod_{k\not=j}^{m} f(\mb_m,\mc_k) 
\prod_{k=1}^{m-1} f(\mb_m,\mb_k)
\mathcal{S}_{\ell,m-1}^{\mod{3}{\mb_m}}
\Big(
\wmb_m, \blb \Big| \b{x}, \wmc_j 
\Big)
\\
\nonumber
&
=
\left(
r_3(z) \prod_{k=1}^{m-1} f(\mb_m,\mb_k)
-
f(\mb_m,\blb) \prod_{k=1}^{m-1} f(\mb_k,\mb_m) 
\right)
\prod_{k\not=j}^{m} f(\mb_m,\mc_k)
\mathcal{S}_{\ell,m-1}^{\mod{3}{\mb_m}}
\Big(
\wmb_m, \blb \Big| \b{x}, \wmc_j 
\Big)
\end{align}
where we obtain the final line of \r{contr1} and \r{contr2} by using the nested Bethe equations \r{bethe-l} and \r{bethe-m} to eliminate $r_1(\lb_{\ell})$ and $r_3(\mb_m)$, respectively. Returning to the formula 
\r{T-action}, we obtain expressions for $\res{z}{\lb_{\ell}} (\mathcal{S}_{\ell,m}(z))$ and $\res{z}{\mb_m} (\mathcal{S}_{\ell,m}(z))$, as sums over terms of the form \r{contr1} and \r{contr2}, respectively. Comparing these sum expressions with their equivalents, \r{act-left-l} and \r{act-left-m}, and cancelling a global common factor, we obtain
\begin{multline}
\label{clean-rec-l}
\mathcal{S}_{\ell,m}( \bmb, \blb | \blc, \bmc )
=
\sum_{i=1}^{\ell} \sum_{j=1}^{m}
g(\mc_j,\lc_i)
\prod_{k\not=i}^{\ell} f(\lc_k,\lb_{\ell})
\prod_{k\not=i}^{\ell} f(\lc_i,\lc_k)
\prod_{k\not=j}^{m} f(\mc_k,\mc_j)
\\
\times
\Big(
g(\mc_j,\lb_{\ell}) \beta_1(\lc_i | \blc, \bmc)
-
g(\lc_i,\lb_{\ell}) \beta_3(\mc_j | \blc, \bmc)
\Big)
\mathcal{S}_{\ell-1,m}^{\mod{1}{\lb_{\ell}}}
\Big(
\bmb, \wlb_{\ell} \Big| \wlc_i, \wmc_j \oplus \lb_{\ell} 
\Big)
\\
+
f(\bmc,\lb_{\ell})
\sum_{i=1}^{\ell}
\prod_{k\not=i}^{\ell} f(\lc_k,\lb_{\ell})
\prod_{k\not=i}^{\ell} f(\lc_i,\lc_k)
g(\lc_i,\lb_{\ell})
\beta_1(\lc_i | \blc, \bmc)
\mathcal{S}_{\ell-1,m}^{\mod{1}{\lb_{\ell}}}
\Big(
\bmb, \wlb_{\ell} \Big| \wlc_i, \bmc 
\Big)
\end{multline}

\begin{multline}
\label{clean-rec-m}
\mathcal{S}_{\ell,m}( \bmb, \blb | \blc, \bmc )
=
\sum_{i=1}^{\ell} \sum_{j=1}^{m}
g(\mc_j,\lc_i)
\prod_{k\not=j}^{m} f(\mb_m,\mc_k)
\prod_{k\not=i}^{\ell} f(\lc_i,\lc_k)
\prod_{k\not=j}^{m} f(\mc_k,\mc_j)
\\
\times
\Big(
g(\mc_j,\mb_m) \beta_1(\lc_i | \blc, \bmc)
-
g(\lc_i,\mb_m) \beta_3(\mc_j | \blc, \bmc)
\Big)
\mathcal{S}_{\ell,m-1}^{\mod{3}{\mb_m}}
\Big( \wmb_m, \blb \Big| \wlc_i \oplus \mb_m, \wmc_j \Big) 
\\
-
f(\mb_m,\blc)
\sum_{j=1}^{m}
\prod_{k\not=j}^{m} f(\mb_m,\mc_k)
\prod_{k\not=j}^{m} f(\mc_k,\mc_j)
g(\mc_j,\mb_m)
\beta_3(\mc_j | \blc, \bmc)
\mathcal{S}_{\ell,m-1}^{\mod{3}{\mb_m}} 
\Big(\wmb_m, \blb \Big| \blc, \wmc_j \Big)
\end{multline}
{\it Important remark.} The scalar products $\mathcal{S}_{\ell-1,m}$ and $\mathcal{S}_{\ell,m-1}$ on the right hand side of \r{clean-rec-l} and \r{clean-rec-m} have their functions $r_1,r_3$ modified according to the rules \r{mod1} and \r{mod3}. In both cases, the nested Bethe equations continue to apply {\it to the modified functions}, since by rearranging \r{bethe-l} and \r{bethe-m} we obtain
\begin{align*}
r_1^{\mod{1}{\lb_{\ell}}}(\lb_i)
\equiv
r_1(\lb_i)
\frac{f(\lb_{\ell},\lb_i)}{f(\lb_i,\lb_{\ell})}
=
- 
\prod_{k=1}^{\ell-1}
\left(
\frac{\lb_k - \lb_i -1}{\lb_k - \lb_i +1}
\right)
\prod_{k=1}^{m}
\left(
\frac{\mb_k - \lb_i +1}{\mb_k - \lb_i}
\right),
&\quad
1 \leq i \leq \ell-1
\\
r_3^{\mod{1}{\lb_{\ell}}}(\mb_j)
\equiv
\frac{r_3(\mb_j)}{f(\mb_j,\lb_{\ell})}
=
- 
\prod_{k=1}^{m}
\left(
\frac{\mb_j - \mb_k -1}{\mb_j - \mb_k +1}
\right)
\prod_{k=1}^{\ell-1}
\left(
\frac{\mb_j - \lb_k +1}{\mb_j - \lb_k}
\right),
&\quad
1 \leq j \leq m
\end{align*}
which apply to \r{clean-rec-l}, and similarly
\begin{align*}
r_1^{\mod{3}{\mb_m}}(\lb_i)
\equiv
\frac{r_1(\lb_i)}{f(\mb_m,\lb_i)}
=
- 
\prod_{k=1}^{\ell}
\left(
\frac{\lb_k - \lb_i -1}{\lb_k - \lb_i +1}
\right)
\prod_{k=1}^{m-1}
\left(
\frac{\mb_k - \lb_i +1}{\mb_k - \lb_i}
\right),
&\quad
1 \leq i \leq \ell
\\
r_3^{\mod{3}{\mb_m}}(\mb_j)
\equiv
r_3(\mb_j)
\frac{f(\mb_j,\mb_m)}{f(\mb_m,\mb_j)}
=
- 
\prod_{k=1}^{m-1}
\left(
\frac{\mb_j - \mb_k -1}{\mb_j - \mb_k +1}
\right)
\prod_{k=1}^{\ell}
\left(
\frac{\mb_j - \lb_k +1}{\mb_j - \lb_k}
\right),
&\quad
1 \leq j \leq m-1
\end{align*}
which apply to \r{clean-rec-m}. Hence all of the scalar products on the right hand side of \r{clean-rec-l} and \r{clean-rec-m} are between on-shell and off-shell states. This is essential, because it allows us to iterate these recursion relations. 

\subsection{Conversion to integral recursion relations}

Equations \r{clean-rec-l} and \r{clean-rec-m} achieve our aim of writing the scalar product 
$\mathcal{S}_{\ell,m}$ explicitly in terms of scalar products $\mathcal{S}_{\ell-1,m}$ and 
$\mathcal{S}_{\ell,m-1}$, respectively. For the purpose of proving \r{mult-int1} and 
\r{mult-int2}, it is useful to convert them into integral recursion relations. For \r{clean-rec-l} we obtain
\begin{multline}
\label{int-rec-l}
\frac{\mathcal{S}_{\ell,m}( \bmb, \blb | \blc, \bmc )}{f(\blc,\blb)}
=
\oint_{\mathcal{X}} \frac{dx}{2\pi i} \oint_{\mathcal{Y}} \frac{dy}{2\pi i}
\frac{
\mathcal{S}_{\ell-1,m}^{\mod{1}{\lb_{\ell}}} 
\Big(\bmb, \blb \ominus \lb_{\ell} \Big| \blc \ominus x, \bmc \oplus \lb_{\ell} \ominus y \Big)
}
{f(\blc \ominus x, \blb \ominus \lb_{\ell})}
\\
\times
g(x,y) g(x,\lb_{\ell}) g(y,\lb_{\ell})
\frac{f(x,\blc)}{f(x,\blb)} f(\bmc,y)
\left(
\frac{\beta_1(x | \blc, \bmc)}
{g(x,\lb_{\ell})}
- 
\frac{\beta_3(y | \blc, \bmc)}
{g(y,\lb_{\ell})}
\right)
\end{multline}
where the contour $\mathcal{X}$ surrounds only poles present at the points $\blc$, and $\mathcal{Y}$ surrounds only poles present at the points $\bmc \oplus \lb_{\ell}$. Due to symmetry in the variables $\blb$, one can write an identical recursion relation with respect to any $\lb_i$, simply by replacing 
$\lb_{\ell} \leftrightarrow \lb_i$ in \r{int-rec-l}. Similarly, by rewriting \r{clean-rec-m} we find
\begin{multline}
\label{int-rec-m}
\frac{\mathcal{S}_{\ell,m}( \bmb, \blb | \blc, \bmc )}{f(\bmb,\bmc)}
=
\oint_{\mathcal{X}} \frac{dx}{2\pi i} \oint_{\mathcal{Y}} \frac{dy}{2\pi i}
\frac{
\mathcal{S}_{\ell,m-1}^{\mod{3}{\mb_m}} 
\Big(\bmb \ominus \mb_m, \blb \Big| \blc \oplus \mb_m \ominus x , \bmc \ominus y \Big)
}
{f(\bmb \ominus \mb_m,\bmc \ominus y)}
\\
\times
g(x,y) g(x,\mb_m) g(y,\mb_m)
f(x,\blc) \frac{f(\bmc,y)}{f(\bmb,y)}
\left(
\frac{\beta_1(x | \blc, \bmc)}
{g(x,\mb_m)}
- 
\frac{\beta_3(y | \blc, \bmc)}
{g(y,\mb_m)}
\right)
\end{multline}
where the contour $\mathcal{X}$ surrounds only poles present at the points $\blc \oplus \mb_m$, and $\mathcal{Y}$ surrounds only poles present at the points $\bmc$. Again, due to symmetry in the variables $\bmb$, we can write an identical relation with respect to any $\mb_j$, by replacing $\mb_m \leftrightarrow \mb_j$ in \r{int-rec-m}.

\subsection{Solution of recursion relations}

For simplicity, we restrict our attention to the solution of just one of the recursion relations, say 
\r{int-rec-m}.\footnote{We will not give the proof that \r{mult-int2} is a solution of \r{int-rec-l}, since it is completely analogous to the proof that we describe in the rest of this section.} We begin by writing this recursion relation in terms of $\mb_1$, purely for convenience:
\begin{multline}
\label{int-rec-m1}
\frac{\mathcal{S}_{\ell,m}( \bmb, \blb | \blc, \bmc )}{f(\bmb,\bmc)}
=
\oint_{\mathcal{X}_1} \frac{dx_1}{2\pi i} \oint_{\mathcal{Y}_1} \frac{dy_1}{2\pi i}
\frac{
\mathcal{S}_{\ell,m-1}^{\mod{3}{\mb_1}} 
\Big(\bmb \ominus \mb_1, \blb \Big| \blc \oplus \mb_1 \ominus x_1 , \bmc \ominus y_1 \Big)
}
{f(\bmb \ominus \mb_1,\bmc \ominus y_1)}
\\
\times
g(x_1,y_1) g(x_1,\mb_1) g(y_1,\mb_1)
f(x_1,\b{X}^1) \frac{f(\b{Y}^1,y_1)}{f(\bmb,y_1)}
\left(
\frac{\beta_1(x_1 | \b{X}^1, \b{Y}^1)}
{g(x_1,\mb_1)}
- 
\frac{\beta_3(y_1 | \b{X}^1, \b{Y}^1)}
{g(y_1,\mb_1)}
\right)
\end{multline}
where $\b{X}^1 = \blc$ and $\b{Y}^1 = \bmc$, and the contour $\mathcal{X}_1$ surrounds only poles present at the points $\blc \oplus \mb_1$, while $\mathcal{Y}_1$ surrounds only poles present at the points $\bmc$. From here, one should iterate a further $m-1$ times until arriving at the base of the recursion. To illustrate the process more clearly, let us write down the next step in the iteration after \r{int-rec-m1}:
\begin{align}
\label{int-rec-m2}
&
\frac{
\mathcal{S}_{\ell,m-1}^{\mod{3}{\mb_1}} 
\Big(\bmb \ominus \mb_1, \blb \Big| \blc \oplus \mb_1 \ominus x_1 , \bmc \ominus y_1 \Big)
}
{f(\bmb \ominus \mb_1,\bmc \ominus y_1)}
=
\\
&
\oint_{\mathcal{X}_2} \frac{dx_2}{2\pi i} \oint_{\mathcal{Y}_2} \frac{dy_2}{2\pi i}
\frac{
\mathcal{S}_{\ell,m-2}^{\mod{3}{\bmb_2}} 
\Big(\bmb \ominus \bmb_2, \blb \Big| \blc \oplus \bmb_2 \ominus \b{x}_2 , \bmc \ominus \b{y}_2 \Big)
}
{f(\bmb \ominus \bmb_2,\bmc \ominus \b{y}_2)}
\nonumber
\\
&
\times
g(x_2,y_2) g(x_2,\mb_2) g(y_2,\mb_2)
f(x_2,\b{X}^2) \frac{f(\b{Y}^2,y_2)}{f(\bmb,y_2)}
\left(
\frac{\beta_1(x_2 | \b{X}^2, \b{Y}^2)}
{g(x_2,\mb_2)}
- 
\frac{\beta_3(y_2 | \b{X}^2, \b{Y}^2)}
{g(y_2,\mb_2)}
\right)
\nonumber
\end{align}
where $\b{X}^2 = \blc \oplus \mb_1 \ominus x_1$ and $\b{Y}^2 = \bmc \oplus \mb_1 \ominus y_1$, and the contour $\mathcal{X}_2$ surrounds only poles present at the points $\blc \oplus \bmb_2 \ominus x_1$, while $\mathcal{Y}_2$ surrounds only poles present at the points $\bmc \ominus y_1$. Some care is needed in writing down \r{int-rec-m2}: it is simply a reiteration of \r{int-rec-m1}, but with the difference that all functions $r_1$ and $r_3$ should be modified according to the rule \r{mod3}. This is crucial to the fact that $\beta_1(x_2|\b{X}^2,\b{Y}^2)$ and $\beta_3(y_2|\b{X}^2,\b{Y}^2)$ appear with precisely the arguments shown.
 
Continuing in this way, we obtain the following solution of the recursion relation:
\begin{multline}
\label{int-sol-m}
\frac{\mathcal{S}_{\ell,m}( \bmb, \blb | \blc, \bmc )}{f(\bmb,\bmc)}
=
\oint_{\mathcal{X}_1} \frac{dx_1}{2\pi i} \oint_{\mathcal{Y}_1} \frac{dy_1}{2\pi i}
\cdots
\oint_{\mathcal{X}_m} \frac{dx_m}{2\pi i} \oint_{\mathcal{Y}_m} \frac{dy_m}{2\pi i}
\mathcal{S}_{\ell,0}^{\mod{3}{\bmb}} 
\Big(\emptyset, \blb \Big| \blc \oplus \bmb \ominus \b{x}, \emptyset \Big)
\\
\times
\prod_{k=1}^{m}
g(x_k,y_k) g(x_k,\mb_k) g(y_k,\mb_k)
f(x_k,\b{X}^k) \frac{f(\b{Y}^k,y_k)}{f(\bmb,y_k)}
\left(
\frac{\beta_1(x_k | \b{X}^k, \b{Y}^k)}
{g(x_k,\mb_k)}
-
\frac{\beta_3(y_k | \b{X}^k, \b{Y}^k)}
{g(y_k,\mb_k)}
\right)
\end{multline}
where the sets $\b{X}^k$ and $\b{Y}^k$ are as given by \r{xy-sets}, and the integration contours 
$\mathcal{X}_k$ and $\mathcal{Y}_k$ surround the points
\begin{align*}
\blc_{\ell} \oplus \bmb_k \ominus \b{x}_{k-1} \subset \mathcal{X}_k,
\quad\quad
\bmc_m \ominus \b{y}_{k-1} \subset \mathcal{Y}_k
\end{align*}
The non-trivial part of the integrand is an $SU(2)$ scalar product of the form \r{su2-sp-l}, but modified such that
\begin{align}
\label{modify}
r_1(z)
\mapsto
r_1^{\mod{3}{\bmb}}(z)
=
\frac{r_1(z)}{f(\bmb,z)}
\quad\quad
\forall\
z \in \blb \oplus \blc \oplus \bmb 
\end{align}
In fact, since the Bethe equations \r{bethe-l} can be written in the form
\begin{align*}
r_1^{\mod{3}{\bmb}}(\lb_i)
=
r_1(\lb_i) 
\prod_{k=1}^{m}
\left(
\frac{\mb_k-\lb_i}{\mb_k-\lb_i+1}
\right)
=
-\prod_{k=1}^{\ell}
\left(
\frac{\lb_k-\lb_i-1}{\lb_k-\lb_i+1}
\right)
\end{align*}
we find that the Slavnov formula \r{slavnov-l} applies to the scalar product present in the integrand of \r{int-sol-m}, but with all functions $r_1(\lc_i), r_1(\mb_i)$ present in the entries of the determinant modified according to \r{modify}. This is what gives rise to a determinant whose entries are of the form \r{entries-mod-1}.

\subsection{Further simplification}

To complete the derivation of \r{mult-int1}, we make some further observations. The first is that the integrand in \r{int-sol-m} does not possess poles at $x_j = \b{x}_{j-1}$ or at $y_j = \b{y}_{j-1}$. Therefore it is not necessary to exclude these points from the integration contours, and we can use the contours as given in \r{contours}. In addition, the order of integration is no longer important after making this change.

The second is that the determinant 
$\mathcal{S}_{\ell,0}^{\mod{3}{\bmb}} 
(\emptyset, \blb | \blc \oplus \bmb \ominus \b{x}, \emptyset )$ should arise as an $\ell \times \ell$ minor of some $(\ell+m) \times (\ell+m)$ determinant, depending on the full set of variables $\blc \oplus \bmb$. To that end, we introduce the determinant \r{ext-slavnov}, which is a Slavnov determinant padded with poles in the $\b{x}$ variables. Paying attention to the slight differences between the integrand of \r{mult-int1} and that of \r{int-sol-m}, it is elementary to show that these two integrations produce the same result. 

\section{Limiting cases of the multiple integral formulae}
\label{s:cases}

As was shown in \cite{whe,fw2}, in the case where a single set of Bethe variables $\blb_{\ell}$ or $\bmb_m$ tends to infinity, the on-shell/off-shell $SU(3)$ scalar product factorizes into a product of two determinants. One of these determinants is a Slavnov determinant for an on-shell/off-shell $SU(2)$ scalar product, while the other is a limiting case thereof. In this section, we recover these results starting from the multiple integral expressions \r{mult-int1} and \r{mult-int2}. 

\subsection{The limit $\bmb_{m} \rightarrow \infty$}
\label{ss:cases-m}

Starting from the expression \r{mult-int1}, we calculate the following limit:
\begin{align*}
\mathcal{S}_{\ell,m}( \infty, \blb, | \blc, \bmc )
\equiv
\frac{1}{m!}
\lim_{\mb_{m},\dots,\mb_1 \rightarrow \infty}
\Big\{
\mb_{m} \dots \mb_1
\mathcal{S}_{\ell,m}( \bmb, \blb | \blc, \bmc )
\Big\}
\end{align*}
Consider the integration over the contours $\mathcal{X}_k$ in \r{mult-int1}, 
$1 \leq k \leq m$, which surround the poles $\blc_{\ell} \oplus \bmb_k$ present in the integrand. In the limit under consideration, only one pole from each contour gives a non-vanishing contribution; namely, the pole at $x_k = \mb_k$, for all $1\leq k \leq m$. Hence we need only retain a single term from the integration over the $m$ contours 
$\mathcal{X}_k$: 
\begin{multline*}
\mathcal{S}_{\ell,m}( \infty, \blb, | \blc, \bmc )
=
\frac{1}{m!}
\lim_{\mb_{m},\dots,\mb_1 \rightarrow \infty}
\left\{
\mb_{m} \dots \mb_1
f(\bmb_m,\blc_{\ell})
\mathcal{S}_{\ell,0}^{\mod{3}{\bmb_m}} 
(\emptyset, \blb | \blc, \emptyset )
\phantom{\prod_{k=1}^{m}}
\right.
\\
\left.
\times 
\oint_{\mathcal{Y}_{m}}
\frac{dy_m}{2\pi\i}
\cdots
\oint_{\mathcal{Y}_1}
\frac{dy_1}{2\pi\i}
\ 
\ivan(\b{y})
(-)^m
\prod_{k=1}^{m}
h(\b{Y}^{k},y_k) 
\beta_3 (y_k | \blc_{\ell},\b{Y}^{k} ) 
g(\bmb_k,y_k)
g(\bmc_m,y_k)
\right\}
\end{multline*}
where $\mathcal{S}_{\ell,0}^{\mod{3}{\bmb_m}} (\emptyset, \blb | \blc, \emptyset)$ denotes a Slavnov determinant \r{slavnov-l}, but whose entries are the functions \r{entries-mod-1}. The integration over the contours $\mathcal{Y}_k$ is unchanged. Continuing with the calculation, we observe that
\begin{multline*}
\mathcal{S}_{\ell,m}( \infty, \blb, | \blc, \bmc )
=
\mathcal{S}_{\ell,0} 
(\emptyset, \blb | \blc, \emptyset )
\times 
\\
\frac{1}{m!}
\lim_{\mb_{m},\dots,\mb_1 \rightarrow \infty}
\left\{
\mb_{m} \dots \mb_1
\oint_{\mathcal{Y}_{m}}
\frac{dy_m}{2\pi\i}
\cdots
\oint_{\mathcal{Y}_1}
\frac{dy_1}{2\pi\i}
\ 
\prod_{k=1}^{m}
g(y_k,\mb_k)
f(\b{Y}^{k},y_k) 
\beta_3 (y_k | \blc_{\ell},\b{Y}^{k} ) 
\right\}
\end{multline*}
where $\mathcal{S}_{\ell,0}(\emptyset, \blb | \blc, \emptyset)$ is the standard Slavnov determinant \r{slavnov-l}, with no modification of the entries. Comparing the multiple integral in this equation with \r{mult-int2-su2} from \s{ss:su2-particular}, we notice just one difference: $\beta_3(y_k,|\blc,\b{Y}^k)$ appears in the integrand, rather than the function 
$\beta_3(y_k,|\emptyset, \b{Y}^k)$ of \r{mult-int2-su2}. This means that the multiple integral evaluates to a modified Slavnov determinant: 
\begin{align*}
&
\frac{1}{m!}
\lim_{\mb_{m},\dots,\mb_1 \rightarrow \infty}
\left\{
\mb_{m} \dots \mb_1
\oint_{\mathcal{Y}_{m}}
\frac{dy_m}{2\pi\i}
\cdots
\oint_{\mathcal{Y}_1}
\frac{dy_1}{2\pi\i}
\ 
\prod_{k=1}^{m}
g(y_k,\mb_k)
f(\b{Y}^{k},y_k) 
\beta_3 (y_k | \blc_{\ell},\b{Y}^{k} ) 
\right\}
\\
=&
\frac{1}{m!}
\lim_{\mb_{m},\dots,\mb_1 \rightarrow \infty}
\left\{
\mb_{m} \dots \mb_1
\mathcal{S}_{0,m}^{\mod{1}{\blc_{\ell}}}
(\bmb, \emptyset | \emptyset, \bmc)
\right\}
\equiv
\mathcal{S}_{0,m}^{\mod{1}{\blc_{\ell}}}
(\infty, \emptyset | \emptyset, \bmc)
\nonumber
\end{align*}
where we have defined
\begin{align*}
\mathcal{S}_{0,m}^{\mod{1}{\blc_{\ell}}}
(\bmb, \emptyset | \emptyset, \bmc)
=
\frac{1}{ \dvan(\bmb) \ivan(\bmc)}
\det\left(
S^{(3)}_j ( \bmb, \blc | \mc_i )
\right)_{1 \leq i,j \leq m}
\end{align*}
Hence we find that, in the limit under consideration, $\mathcal{S}_{\ell,m}$ exhibits the factorization
\begin{align}
\mathcal{S}_{\ell,m}( \infty, \blb, | \blc, \bmc )
=
\mathcal{S}_{\ell,0} 
(\emptyset, \blb | \blc, \emptyset )
\times
\mathcal{S}_{0,m}^{\mod{1}{\blc}}
(\infty, \emptyset | \emptyset, \bmc )
\end{align}
which agrees with the result found in \cite{whe,fw2}, up to normalization.

\subsection{The limit $\blb_{\ell} \rightarrow \infty$}

Starting this time from the expression \r{mult-int2}, we calculate the limit
\begin{align*}
\mathcal{S}_{\ell,m}( \bmb, \infty | \blc, \bmc )
\equiv
\frac{1}{\ell!}
\lim_{\lb_{\ell},\dots,\lb_1 \rightarrow \infty}
\Big\{
\lb_{\ell} \dots \lb_1
\mathcal{S}_{\ell,m}( \bmb, \blb | \blc, \bmc )
\Big\}
\end{align*}
The procedure for this calculation directly parallels that of \s{ss:cases-m}, so we only sketch the details. In the limit under consideration, only one pole from each contour $\mathcal{Y}_k$ gives a non-vanishing contribution; namely, the pole at $y_k = \lb_k$, for all 
$1 \leq k \leq \ell$. Hence it suffices to retain just one term from this integration:
\begin{multline*}
\mathcal{S}_{\ell,m}( \bmb, \infty | \blc, \bmc )
=
\frac{1}{\ell!}
\lim_{\lb_{\ell},\dots,\lb_1 \rightarrow \infty}
\left\{
\lb_{\ell} \dots \lb_1
f(\bmc_m,\blb_{\ell})
\mathcal{S}_{0,m}^{\mod{1}{\blb_{\ell}}} 
(\bmb, \emptyset | \emptyset, \bmc )
\phantom{\prod_{k=1}^{\ell}}
\right.
\\
\left.
\times 
\oint_{\mathcal{X}_{\ell}}
\frac{dx_{\ell}}{2\pi\i}
\cdots
\oint_{\mathcal{X}_1}
\frac{dx_1}{2\pi\i}
\ 
\dvan(\b{x})
(-)^{\ell}
\prod_{k=1}^{\ell}
h(x_k,\b{X}^{k}) 
\beta_1 (x_k | \b{X}^k, \bmc_m ) 
g(x_k,\blb_k)
g(x_k,\blc_{\ell})
\right\}
\end{multline*}
where $\mathcal{S}_{0,m}^{\mod{1}{\blb_{\ell}}} (\bmb, \emptyset | \emptyset, \bmc )$ denotes a Slavnov determinant \r{slavnov-m}, but whose entries are the modified functions \r{entries-mod-3}. Clearly, we can simplify this further:
\begin{multline*}
\mathcal{S}_{\ell,m}( \bmb, \infty | \blc, \bmc )
=
\mathcal{S}_{0,m} 
( \bmb, \emptyset | \emptyset, \bmc )
\times 
\\
\frac{1}{\ell!}
\lim_{\lb_{\ell},\dots,\lb_1 \rightarrow \infty}
\left\{
\lb_{\ell} \dots \lb_1
\oint_{\mathcal{X}_{\ell}}
\frac{dx_{\ell}}{2\pi\i}
\cdots
\oint_{\mathcal{X}_1}
\frac{dx_1}{2\pi\i}
\ 
\prod_{k=1}^{\ell}
g(\lb_k,x_k)
f(x_k,\b{X}^{k}) 
\beta_1 (x_k | \b{X}^k, \bmc_m) 
\right\}
\end{multline*}
where $\mathcal{S}_{0,m} ( \bmb, \emptyset | \emptyset, \bmc )$ is now the standard Slavnov determinant \r{slavnov-m}. Comparing with \r{mult-int1-su2} from 
\s{ss:su2-particular}, we conclude that
\begin{align*}
&
\frac{1}{\ell!}
\lim_{\lb_{\ell},\dots,\lb_1 \rightarrow \infty}
\left\{
\lb_{\ell} \dots \lb_1
\oint_{\mathcal{X}_{\ell}}
\frac{dx_{\ell}}{2\pi\i}
\cdots
\oint_{\mathcal{X}_1}
\frac{dx_1}{2\pi\i}
\ 
\prod_{k=1}^{\ell}
g(\lb_k,x_k)
f(x_k,\b{X}^{k}) 
\beta_1 (x_k | \b{X}^k, \bmc_m) 
\right\}
\\
=&
\frac{1}{\ell!}
\lim_{\lb_{\ell},\dots,\lb_1 \rightarrow \infty}
\Big\{
\lb_{\ell} \dots \lb_1
\mathcal{S}_{\ell,0}^{\mod{3}{\bmc_m}}
( \emptyset, \blb | \blc, \emptyset )
\Big\}
\equiv
\mathcal{S}_{\ell,0}^{\mod{3}{\bmc_m}}
(\emptyset, \infty | \blc, \emptyset )
\nonumber
\end{align*}
where we have defined the modified Slavnov determinant
\begin{align*}
\mathcal{S}_{\ell,0}^{\mod{3}{\bmc_m}} (\emptyset, \blb | \blc, \emptyset)
&=
\frac{1}{ \ivan(\blb) \dvan(\blc) }
\det\left(
S^{(1)}_j ( \bmc, \blb | \lc_i )
\right)_{1 \leq i,j \leq \ell}
\end{align*}
Therefore, in this limit we have the factorization
\begin{align}
\mathcal{S}_{\ell,m}( \bmb, \infty | \blc, \bmc )
=
\mathcal{S}_{0,m} 
( \bmb, \emptyset | \emptyset, \bmc )
\times
\mathcal{S}_{\ell,0}^{\mod{3}{\bmc}}
(\emptyset, \infty | \blc, \emptyset )
\end{align}
which agrees, once again, with the result obtained in \cite{whe,fw2} up to normalization.

\section{Discussion}
\label{s:discuss}

The formulae \r{mult-int1} and \r{mult-int2} which we have presented in this paper are by no means the first examples of multiple integral expressions for scalar products, or related objects. In the case of the $SU(2)$-invariant XXX spin-1/2 Heisenberg chain, different multiple integral expressions for the scalar product were found in \cite{gal1,dgs} and more recently, using the method of separation of variables due to Sklyanin, in \cite{kkn}. Further to this, in generic $SU(2)$-invariant models, the {\it master equation} of \cite{kkmst} expresses a generating series of correlation functions as a multiple integral. Much like the formulae in this paper, the integrand appearing in \cite{kkmst} depends on Slavnov determinants and the contours surround poles at Bethe roots. We mention also \cite{bpr}, where the scalar product of models based on $U_q(\widehat{sl_3})$ (the trigonometric generalization of the rational model studied here) was expressed as a certain multiple integral. However, in contrast with the results \r{mult-int1} and \r{mult-int2} in the present paper, the integrand appearing in \cite{bpr} has a particularly complicated form and is not expressed as a determinant. 

We expect that the multiple integral formulae in this work can have application in two directions. Firstly, as we mentioned already in the introduction, by virtue of the results in \cite{bprs3} it is possible to write an arbitrary correlation function in an $SU(3)$-invariant model as a sum over on-shell/off-shell scalar products. With the help of equations \r{mult-int1} and \r{mult-int2} one should obtain, in principle, an exact and manageable expression for such correlation functions in finite size. The hope is that these expressions will be sufficiently simple to permit the study of asymptotics. 

Secondly, the scalar product of the $SU(2)$-invariant XXX spin-1/2 chain plays a key role in calculating the tree-level correlation function of three single-trace operators in $su(2)$ sectors, in planar $\mathcal{N}=4$ super Yang-Mills \cite{egsv,fod}. Recently in \cite{fjks}, the authors considered an extension of this calculation to the case of one $su(2)$ and two $su(3)$ operators, and found that the resulting 3-point function can be expressed in terms of scalar products of $SU(3)$ Bethe vectors. With the help of the multiple integral formulae in this paper, one thus obtains an exact expression for these 3-point functions in finite size. An important step in the study of asymptotics would be the calculation of the semi-classical limit of \r{mult-int1} and \r{mult-int2}, generalizing the work of Kostov in \cite{kos1,kos2}, where the semi-classical limit of the Slavnov determinant was evaluated. While such a calculation is beyond the scope of this work, we plan to study this further.

Finally, let us remark that it would be worthwhile to understand how the expressions \r{mult-int1} and \r{mult-int2} reduce to single determinants, when the free sets of variables $\blc$ and $\bmc$ are set equal to {\bf 1.} The sets $\blb$ and $\bmb$, respectively, or {\bf 2.} Roots of the twisted Bethe equations (corresponding to a twisted transfer matrix). In case {\bf 1}, which is a singular limit of \r{mult-int1} and \r{mult-int2}, we expect to recover the determinant formula for the norm-squared, obtained in \cite{res}. In case {\bf 2}, we should recover the determinant found in \cite{bprs2}. Understanding these limits properly is important, since this may well lead to further simplifications of the formulae \r{mult-int1} and \r{mult-int2} themselves.

\end{document}